\documentclass[prd,superscriptaddress,amsfonts,amssymb,amsmath,showpacs,twocolumn]{revtex4-2}
\usepackage{bm}
\usepackage{amsfonts}
\usepackage{latexsym}
\usepackage[latin1]{inputenc}
\usepackage{graphicx}
\usepackage{amsmath,eqnarray}
\usepackage{palatino}
\usepackage{mathpazo}
\usepackage{textcomp}
\usepackage{float}
\usepackage{booktabs}
\usepackage{dcolumn}
\usepackage{hyperref}
\usepackage{mathtools}
\hypersetup{colorlinks,citecolor=red}

\usepackage{amsmath}
\usepackage{xcolor}
\usepackage{orcidlink}
\usepackage[caption=false]{subfig}
\usepackage{commath}
\def\jnl@style{\it}
\def\aaref@jnl#1{{\jnl@style#1}}

\def\aaref@jnl#1{{\jnl@style#1}}

\def\aj{\aaref@jnl{AJ}}                   
\def\apj{\aaref@jnl{ApJ}}                 
\def\apjl{\aaref@jnl{ApJ}}                
\def\apjs{\aaref@jnl{ApJS}}               
\def\apss{\aaref@jnl{Ap\&SS}}             
\def\aap{\aaref@jnl{A\&A}}                
\def\aapr{\aaref@jnl{A\&A~Rev.}}          
\def\aaps{\aaref@jnl{A\&AS}}              
\def\mnras{\aaref@jnl{Mon.~Not.~Roy.~Astron.~Soc.}}             
\def\prd{\aaref@jnl{Phys.~Rev.~D}}        
\def\prc{\aaref@jnl{Phys.~Rev.~C}}  
\def\prl{\aaref@jnl{Phys.~Rev.~Lett.}}    
\def\qjras{\aaref@jnl{QJRAS}}             
\def\skytel{\aaref@jnl{S\&T}}             
\def\ssr{\aaref@jnl{Space~Sci.~Rev.}}     
\def\zap{\aaref@jnl{ZAp}}                 
\def\nat{\aaref@jnl{Nature}}              
\def\aplett{\aaref@jnl{Astrophys.~Lett.}} 
\def\apspr{\aaref@jnl{Astrophys.~Space~Phys.~Res.}} 
\def\physrep{\aaref@jnl{Phys.~Rep.}}      
\def\physscr{\aaref@jnl{Phys.~Scr}}       
\def\commat{\aaref@jnl{Comm.~Math.~Phys.}}              
\def\science{\aaref@jnl{Science}}               
\def\cqg{\aaref@jnl{Classical Quant.~Grav.}}            
\def\jpcs{\aaref@jnl{JPCS}}                                     
\def\ijmpd{\aaref@jnl{Int.~J.~Mod.~Phys.~D}}                    
\def\grg{\aaref@jnl{Gen.~Relat.~Gravit.}}               
\def\rpp{\aaref@jnl{Rep.~Prog.~Phys.}}          
\def\npa{\aaref@jnl{Nucl.~Phys.~A}}        
\def\lrr{\aaref@jnl{Living Rev.~Rel.}}                   
\def\jcap{\aaref@jnl{J.~Cosmology Astropart.~Phys.}}    
\def\rmp{\aaref@jnl{Rev.~Mod.~Phys.}}   
\def\epjc{\aaref@jnl{Eur.~Phys.~J.~C}}

\allowdisplaybreaks[1]

\begin{document}

\color{black}

\title{The Stability of Anisotropic Compact Stars Influenced by Dark Matter under Teleparallel Gravity: An Extended Gravitational Deformation Approach}

\author{Sneha Pradhan\orcidlink{0000-0002-3223-4085}}
\email{snehapradhan2211@gmail.com}
\affiliation{Department of Mathematics, Birla Institute of Technology and
Science-Pilani,\\ Hyderabad Campus, Hyderabad-500078, India.}

\author{Piyali Bhar\orcidlink{0000-0001-9747-1009}}
\email{piyalibhar90@gmail.com}
\affiliation{Department of Mathematics, Government General Degree College Singur, Hooghly,
 West Bengal 712409, India}

\author{Sanjay Mandal\orcidlink{0000-0003-2570-2335}}
\email{sanjaymandal960@gmail.com}
\affiliation{Faculty of Symbiotic Systems Science, Fukushima University, Fukushima 960-1296, Japan}

\author{P.K. Sahoo\orcidlink{0000-0003-2130-8832}}
\email{pksahoo@hyderabad.bits-pilani.ac.in}
\affiliation{Department of Mathematics, Birla Institute of Technology and
Science-Pilani,\\ Hyderabad Campus, Hyderabad-500078, India.}

\author{Kazuharu Bamba \orcidlink{0000-0001-9720-8817}}
\email{bamba@sss.fukushima-u.ac.jp}
\affiliation{Faculty of Symbiotic Systems Science, Fukushima University, Fukushima 960-1296, Japan}

\date{\today}
\begin{abstract}
In our investigation, we pioneer the development of geometrically deformed strange stars within the framework of teleparallel gravity theory through gravitational decoupling via the complete geometric deformation (CGD) technique. The significant finding is the precise solution for deformed strange star (SS) models achieved through the vanishing complexity factor scenario. Further, we introduce the concept of space-time deformation caused by dark matter (DM) content in DM haloes, leading to perturbations in the metric potentials $g_{tt}$ and $g_{rr}$ components. Mathematically, this DM-induced deformation is achieved through the CGD method, where the decoupling parameter $\alpha$ governs the extent of DM influence. To validate our findings, we compare our model predictions with observational constraints, including GW190814 (with a mass range of $2.5-2.67 M_{\odot}$) and neutron stars (NSTRs) such as EXO 1785-248 [mass=$1.3_{-0.2}^{+0.2}~M_{\odot}$], 4U 1608-52 [mass=$1.74_{-0.14}^{+0.14}~M_{\odot}$], and PSR J0952-0607 [mass=$2.35_{-0.17}^{+0.17}~M_{\odot}$]. 
 Our investigation delves into the stability of the model by considering causality conditions, Herrera's Cracking Method, the adiabatic index, and the Harrison-Zeldovich-Novikov criterion. We demonstrate that the developed model mimics a wide range of recently observed pulsars. To emphasize its compatibility, we highlight the predicted mass and radius in tabular form by varying both the parameters $\alpha$ and $\zeta_1$. Notably, our findings are consistent with the observation of gravitational waves from the first binary merger event. Furthermore, we compare our results with those obtained for a slow-rotating configuration. In addition to this, we discuss the moment of inertia using the Bejger-Haensel approach in this formulation.


\end{abstract}

\maketitle

\section{Introduction}

Over the past few decades, there has been a significant surge in interest in the topic-modified theories of gravitation. It is widely recognized that with the inclusion of additional matter fields, such as the inflaton, the General Theory of Relativity (GR) can effectively describe the evolution of the universe from its early stages to its current accelerated expansion. However, despite this, GR struggles to account for recent cosmological observations and fails to address various cosmological issues, such as the problems of dark energy, dark matter, and the Hubble tension, among others. Additionally, GR is a non-renormalizable theory and encounters divergences with higher-loop contributions. In this context, the aforementioned modified theories of gravitation could provide significant assistance.
These modified theories of gravitation are typically characterized by an altered Lagrangian density (excluding non-Lagrangian theories like MOND), which is modified by introducing additional geometrodynamical terms into the Einstein-Hilbert action integral. There are several modified theories of gravitation that have been proposed to address the limitations of GR $f (\mathcal{R})$ \cite{1,2,3,4,5}, $f (\mathcal{R}, \mathcal{G})$ \cite{6,7}, $f (\mathcal{G})$ \cite{8}, $f (\mathcal{R}, T )$ \cite{9,10}, $f (\mathcal{Q})$ \cite{11}, $f (\mathcal{R}, \phi)$ \cite{12}, and $f (\mathcal{R}, \phi, X)$ \cite{13}. In the work referenced as \cite{kb1}, the author provides a comprehensive review of the various applications of $f(\mathcal{R})$ theories in cosmology and gravity. This includes discussions on inflation, dark energy, local gravity constraints, cosmological perturbations, and spherically symmetric solutions in weak and strong gravitational fields.
In the article mentioned above, referenced as \cite{5}, the author presents the formalism of several standard modified gravity theories, such as $f(\mathcal{R})$, $f(\mathcal{G})$, and $f(\mathcal{T})$ gravity theories. It also explores various alternative theoretical proposals that have emerged in the literature over the past decade. The article emphasizes the formalism developed for these theories and explains how they can be viable descriptions of our Universe.
A detailed study on teleparallel gravity, ranging from its theoretical foundations to its cosmological implications, is presented in the insightful articles \cite{kb3} and \cite{kb4}

An alternative approach involves starting with the torsional formulation of gravity, specifically the Teleparallel Equivalent of General Relativity (TEGR) \cite{ft11,ft2}, and introducing modifications accordingly. This leads to theories such as $f(\mathcal{T})$ gravity\cite{ft33} , $f(\mathcal{T}, T_G)$ gravity \cite{ft4}, $f(\mathcal{T}, B)$ gravity \cite{ft5}, and scalar-torsion theories\cite{ft6}. On the other hand, extending the gravitational action by introducing terms of the form $f(\mathcal{R})$ or $f(\mathcal{T})$, where $f$ is a nonlinear function, results in significant differences between curvature-based and torsion-based theories of gravity. The extensively studied $f(\mathcal{R})$ gravity, which relies on curvature, has been thoroughly examined over the past decades \cite{ft7}. A notable aspect of this theory is that the resulting equations of motion involve fourth-order derivatives of the metric, which is a distinguishing feature. In sharp contrast, the equations of motion in the torsion-based $f(\mathcal{T})$ theory consist solely of the standard second-order derivatives of the tetrad fields, with no higher-order derivatives. The properties of the teleparallel theory have been extensively explored in the cosmological domain, as evidenced by numerous studies. Constraints on the theory have also been derived by analyzing the motion of planets in the Solar System. In the paper \cite{new1}, the author explores scenarios in which the standard assumption of a strictly positive effective dark energy (DE) density, an assumption commonly held within general relativity frameworks may not necessarily apply. This relaxation opens up novel cosmological possibilities. Specifically, by concentrating on an exponential infrared 
$f(\mathcal{T})$ model, the study investigates the implications of such modifications for cosmological models, expanding the scope for alternative interpretations and behaviors in dark energy dynamics.

However, The black holes observation seeks a lot of interest in the study of compact objects. And, there after many studies have been done exploring the compact objects. In this view, the concept of teleparallel gravity is used to discuss compact stellar objects including black holes, neutron stars, and strange stars in references \cite{fts1}-\cite{ftc5}. Researchers \cite{ftc6} provide details on isotropic compact stars that are described in the setting of $f(\mathcal{T})$ gravity using both diagonal and off-diagonal tetrads. Furthermore, gravitational waves and their characteristics were investigated by Farrugia et al.\cite{fts5} using a variety of modified teleparallel theories. In addition to the teleparallel gravity, several studies on the compact star are available in the literature \cite{fr1}-\cite{fr5}. When modeling realistic, compact objects, it is crucial to account for the effects of local anisotropy $(p_t \neq p_r)$. Typically, the differences in pressures along radial and transverse directions create this anisotropy within the system. New investigations of compact star objects such as PSRJ 1614-2230, 4U 1820-30, and 4U 1538-52 have demonstrated the presence of anisotropic matter distributions within the interiors of these stellar formations. Pioneering work by Bowers and Liang \cite{sd1} on anisotropic compact stars opened up a new area of astrophysical research. Subsequently, many researchers delved into anisotropic stellar systems and their implications. Herrera and Santos \cite{sd2} extensively explored the origin and impact of local anisotropy on the physical properties of compact stars.
Anisotropy in stellar structures can arise from various sources, including a mixture of different fluid types \cite{sd3}, the presence of superfluids or solid cores \cite{sd4}, phase transitions \cite{sd5}, and pion condensation \cite{sd6}. Additionally, strong magnetic fields (commonly known as magnetars) and intense electromagnetic fields \cite{sd7} can also induce anisotropy. In such cases, anisotropic effects are expected due to tensions within the stellar interior \cite{sd8}-\cite{sd12}.
Furthermore, the nature of anisotropy leads to significant changes in stellar structure \cite{sd13}-\cite{sd18}. Notably, anisotropy plays a role in explaining large redshifts and contributes to stellar stability.

Utilizing the groundbreaking concepts introduced by Herrera \cite{sd19}, Abreu et al. \cite{sd20} investigated the stability of compact stars in the presence of density fluctuations and local anisotropy. Their findings indicate that a stellar system can achieve potential stability if the condition $-1 \leq v_t^2 - v_r^2 \leq 0$ is satisfied, while potential instability arises when $ 0 \leq v_t^2 - v_r^2 \leq 1 $. Notably, mere density perturbations alone are insufficient to disturb equilibrium; both density and anisotropy perturbations play essential roles \cite{sd19},\cite{sd21}. In contemporary astrophysical research, anisotropy within modified gravity theories has garnered attention, particularly in the study of wormholes and compact stars \cite{sd22}-\cite{sd25}. Researchers have employed various techniques to integrate the field equations, including conformal motions, embedding class one, conformally flat spacetime, ansatz metric potentials, and equation of states. Among these methods, three stand out as comparatively simpler: conformal motions, embedding class one, and conformally flat spacetime. In each of these approaches, the metric potentials $g_{tt}$
 and $g_{rr}$ exhibit interconnected behavior.

In our study of compact stars under teleparallel gravity, we employ the gravitational decoupling (GD) technique to simplify the field equations. Gravitational decoupling plays a pivotal role in understanding the behavior of compact objects and cosmological solutions, discovered by \cite{gd1}-\cite{gd3}.  There are two significant approaches: the Minimal Geometric Deformation (MGD) technique\cite{gd2} and the Complete Geometric Deformation (CGD) approach \cite{gd1}. The MGD approach has been widely used in GR to generate exact anisotropic solutions from perfect fluid seed solutions. It allows us to decouple gravitational sources and explore more realistic scenarios. Notably, it has been applied to derive spherically symmetric solutions of the Einstein-scalar system from the Schwarzschild vacuum metric \cite{gd4}. The MGD method has been used to generate exact spherically symmetric solutions in the context of fourth order Einstein Weyl gravity. By applying the MGD technique to the Schwarzschild metric, scientists obtain hairy black hole solutions expressed as series using the Homotopy Analysis Method (HAM) \cite{gd4}. The CGD method extends the concept of gravitational decoupling to more complex scenarios. By relaxing assumptions about perfect fluid sources, we can investigate static and spherically symmetric solutions with sources that better represent physical reality. For instance, the CGD approach modifies well-known cosmological solutions\cite{gd6}. Additionally, it influences the complexity factor introduced by Herrera in self-gravitating systems\cite{gd7}.
 In classical GR, the combination of vanishing complexity in self-gravitating stellar systems, anisotropy, and GD provides a suitable set of solutions. The complexity factor concept in the context of GD was introduced by \cite{gd8} and extensively applied by Herrera et al. (\cite{gd9},\cite{gd10}) in spherically symmetric static systems under various physical conditions. In this study, we introduced a straightforward and efficient method for generating new solutions for self-gravitating objects through the CGD approach. Furthermore, our objective is to derive a physically viable solution using the same technique, applied to compact stars with dark matter halos. Dark matter (DM) halos constitute structures composed primarily of dark matter, serves as the gravitational framework for galaxies and galaxy clusters. DM particles, which sufficiently weakly interact undergo gravitational capture and accumulate within compact stars, influence their structure and thermal evolution \cite{dn1}. The Navarro-Frenk-White (NFW) profile remains a widely adopted model to describe the density distribution of dark matter within halos. Derived from numerical simulations of structure formation, it provides an excellent fit to the density profiles of simulated dark matter halos over a broad mass range \cite{dn4}. The NFW profile, or similar density distributions, often models dark matter around astrophysical objects; however, in compact stars, self-gravitating dark matter forms a dense core that significantly impacts the equation of state of the stellar interior \cite{dn2}. A recent study considers three interacting dark energy (IDE) models and measures the allowed limits of IDE coupling parameters for individual cases \cite{dn3}. Our analysis reveals that the presence of dark matter within the stellar fluid plays a significant role in the formation of more massive stars. This finding is consistent with prior theoretical expectations, as the interaction between dark matter and the ordinary matter constituting compact stars strengthens the EoS, making it stiffer. A stiffer EoS enables the star to support a higher mass and achieve greater compactness, ultimately resulting in more massive stellar objects. In addition to this, dark matter influences not only the interactions between itself and ordinary matter but also reinforces the interactions among the particles of ordinary matter. The interplay between dark matter and ordinary matter is thus not merely a passive one. Rather, it introduces an additional layer of stability that allows the star to sustain higher masses and maintain a more orderly structure. This ``gluing" effect, by stabilizing the system, could also influence the star's evolution, leading to more stable and possibly longer-lived stellar bodies in regions where dark matter is present in significant quantities.\\ 
The outline of our paper is organized as follows:
In Section \ref{sec2}, we provide a concise geometrical overview of the field equations in teleparallel gravity, utilizing the CGD technique. Section \ref{seciii} details the derivation of an exact solution for the anisotropic dark star model by employing the vanishing complexity-free approach for both the seed system and the $\Theta$ sector. Sections \ref{seciv} and \ref{secv} present a summary of the physical and stability analyses, validating our constructed model through Herrera's Cracking Method, the adiabatic index, and the Harrison-Zeldovich-Novikov criterion. In Sections \ref{secvi} and \ref{secvii}, we analyze the measurement of maximum mass and radii of observed compact objects using M-R curves and equi-mass diagrams. Section \ref{secviii} addresses the energy exchange between fluid distributions under dark matter. In Section \ref{secix}, we compare our results with a slow-rotating configuration, with a particular focus on the moment of inertia using the Bejger-Haensel approach. Furthermore, a comparative study has been done with the recent advancement on compact stars in Section-\ref{sec10}. Finally, we conclude our investigation by highlighting the key findings in Section \ref{secx}.

\section{Field equation in teleparallel gravity under CGD background}\label{sec2}

This section represents the geometrical background of teleparallel gravity and the equation of motion under the influence of the complete geometric deformation circumstance. For the sake of understanding, let us mention that, Greek indices are used to denote the coordinates of the space-time manifold in this convention. In contrast, Latin indices are reserved for the components of the tangent space associated with the manifold (space-time). In teleparallelism, a vierbein field \( e_i(x^{\mu}) \) for \( i = 0, 1, 2, 3 \) is employed as the dynamical object. This field provides an orthonormal basis for the tangent space at each point \( x^{\mu} \) within the manifold. Each vector \( e_i \) can be described in a coordinate basis by its components \( e_i^{\mu} \), i.e., \( e_i = e_i^{\mu} \partial_{\mu} \) where \( \mu = 0, 1, 2, 3 \). For any given spacetime metric, the line element can be expressed as:
\[ ds^2 = g_{\mu\nu} dx^{\mu} dx^{\nu} = \eta_{ij} e^i_{\mu} e^j_{\nu} dx^{\mu} dx^{\nu} \]
Here, \(\eta_{ij} = diag(-1, +1, +1, +1)\)/$diag(+1, -1, -1, -1)$ (depends on the convention) represents the Minkowski metric. Unlike GR, which is based on the torsion-free Levi-Civita connection, teleparallelism uses the Weitzenb$\ddot{\text{o}}$ck connection :
\[ \widehat{\gamma}^{\lambda}_{\,\,\,\nu\mu} = e^{\lambda}_{\,\,A} \partial_{\mu} e^A_{\,\,\,\nu} \]
as reviewed by \cite{ft1}. This connection defines a geometry that is torsion-based but free of curvature. The torsion tensor can be defined as follows:
\begin{eqnarray}
\mathcal{T}^{\lambda}_{\,\,\,\,\,\mu\nu} = \widehat{\gamma}^{\lambda}_{\,\,\,\nu\mu}-\widehat{\gamma}^{\lambda}_{\,\,\,\mu\nu} = e^{\lambda}_{A}\big(\partial_{\mu}e^A_{\,\,\,\nu}-\partial_{\nu}e^A_{~\mu}\big).
\end{eqnarray}
In numerous calculations, torsion frequently appears in linear combinations, particularly in the contortion tensor, which is defined as follows:
\begin{eqnarray}
\mathcal{K}^{\mu\nu}_{~~~\rho} = -\frac{1}{2}\big(\mathcal{T}^{\mu\nu}_{~~~\rho}-\mathcal{T}^{\nu\mu}_{~~~~\rho}-\mathcal{T}_{\rho}^{~\mu\nu}\big).
\end{eqnarray}
In teleparallel gravity, the well known super-potential $\mathcal{S}_{\rho}^{~~\mu\nu}$ could be presented as :
\begin{eqnarray}
    \mathcal{S}_{\rho}^{~~\mu\nu} = \frac{1}{2}\Big(\mathcal{K}^{\mu\nu}_{~~~~\rho}+\,\delta^{\mu}_{\rho}~\mathcal{T}^{\alpha\nu}_{~~~~\alpha}-\delta^{\nu}_{\rho}~\mathcal{T}^{\alpha\mu}_{~~~~\alpha}\Big).
\end{eqnarray}
Then the teleparallel Lagrangian, defined as torsion scalar, can be expressed using these quantities as follows: \cite{ft3}:
\begin{eqnarray}
&&\hspace{0cm}\mathcal{T}= \mathcal{S}_{\rho}^{~~~\mu\nu}~\mathcal{T}^{\rho}_{~~~\mu\nu}\nonumber\\&&\hspace{0.5cm}=\frac{1}{4}\Big(\mathcal{T}^{\rho\mu\nu}\mathcal{T}_{\rho\mu\nu}+2\mathcal{T}^{\rho\mu\nu}\mathcal{T}_{\nu\mu\rho}-4\mathcal{T}_{~~~\rho\mu}^{\rho}\mathcal{T}^{\nu\mu}_{~~~\nu}\Big).
\end{eqnarray}
In TEGR, the torsion tensor \(\mathcal{T}^{\lambda}_{~~~\mu\nu}\) contains all information about the gravitational sector, much like how the Riemann curvature tensor gives rise to the curvature scalar \(\mathcal{R}\) in Einstein's GR. 
In teleparallel gravity, the action is formulated using the torsion scalar \(\mathcal{T}\). The concept of \(f(\mathcal{T})\) gravity involves extending \(\mathcal{T}\) into a function \(f(\mathcal{T})\). This is analogous to the approach $f(\mathcal{R})$ gravity from GR. 
The E-H action can be expressed as :
\begin{eqnarray}\label{action}
    \mathcal{S}=\int \sqrt{-g} \Big[\frac{1}{16\pi }f(\mathcal{T})+\mathcal{L_{\mathrm{M}}}+\mathcal{L}_{\Theta}\Big]d^4x.
\end{eqnarray}
In the above expression, \( g = \text{det}(g_{\mu\nu}) \). The term \(\mathcal{L}_{\mathrm{M}}\) is known as Lagrangian density for the matter fields, which is associated with the stress-energy tensor \( T_{\mu\nu} \). Meanwhile, \(\mathcal{L}_{\Theta} \) denotes the Lagrangian density related to the new gravitational sector, generally referred to as the "\(\Theta\) gravitational sector" \(\Theta_{\mu\nu}\). This additional gravitational component consistently influences the dark matter fields within \( f(\mathcal{T}) \) gravity and can be incorporated into the total energy-momentum tensor as \( T_{\mu\nu}^{\text{tot}} = T_{\mu\nu} + \alpha \Theta_{\mu\nu} \). Here, \(\alpha\) is the coupling constant that defines the interaction between the matter fields and the \(\Theta\) gravitational sector. By varying the action for the vierbein fields, one can obtain the equation of motion in teleparallel gravity as follows:
\begin{eqnarray}\label{field}
& e_A^{\rho} ~S_{\rho}^{~~\mu\nu} \partial_{\mu}(\mathcal{T}) f_{\mathcal{T} \mathcal{T}}+e^{-1}\partial_{\mu}(e e_A^{\rho} S_{\rho}^{~~\mu \nu}) f_{\mathcal{T}}-e_A^{\lambda} \mathcal{T}_{~~\mu\lambda }^{\rho} \nonumber\\&&\hspace{-8cm} \times S_{\rho}^{~~\nu\mu} f_{\mathcal{T}}+\frac{1}{4} e^{\nu}_{A} f(\mathcal{T})  =  4 \pi e_{\rho}^A\left\{T_{\rho}^{~~\nu}\right\}^{\mathrm{tot}}.
\end{eqnarray}
Where, $\big\{T_{\rho}^{~\nu}\big\}^{\text{tot}}=T_{\rho}^{~\nu}+ \Theta_{\rho}^{~\nu}$, $f_{\mathcal{T}}=\frac{\partial f}{\partial \mathcal{T}}$ and $f_{\mathcal{T}\mathcal{T}}=\frac{\partial^{2} f}{\partial \mathcal{T}^2}$.
The total energy-momentum tensor for the anisotropic fluid matter can be described as follows:
\begin{eqnarray}
\big\{T_{\rho}^{~~\nu}\big\}^{\mathrm{tot}}=(\rho^{\text{tot}}+p_t^{\text{tot}}) U_{\nu} U^{\rho}-p_t^{\text{tot}} \delta_{\nu}^{\rho}+
(p_r^{\text{tot}}-p_t^{\text{tot}}) V_{\nu} V^{\rho}
~~~~
\end{eqnarray}

In the above expression, \(\rho^{\text{tot}}\) denotes the total energy density, where \(p_r^{\text{tot}}\) indicates the total pressure of the fluid towards the $r-r$ component relative to the time-like four-velocity vector \(U_{\nu}\) (known as radial pressure), and \(p_t^{\text{tot}}\) signifies the pressure orthogonal to \(U^{\rho}\) (reffered as tangential pressure). The vector \(U_{\nu}\) represents the four-velocity along a time-like direction, where the term \(V_{\nu}\) denotes the unit space-like vector towards the $r-r$ coordinate direction. Consequently, we describe the dense matter by the anisotropic perfect fluid, where the components of the energy-momentum tensor for the matter with the $\Theta$ gravitational sector could be written as \(\big(-\rho^{\text{tot}}, p_r^{\text{tot}}, p_t^{\text{tot}}, p_t^{\text{tot}}\big)\). The mathematical expression of the energy-momentum tensor can be expressed as follows:
\begin{eqnarray}
    \hspace{0.5cm}\rho^{\text{tot}}=\rho+\alpha \big[T^{\Theta}\big]^{0}_{0}\,, \\
   \hspace{0.5cm} p_r^{\text{tot}}=p_r-\alpha \big[T^{\Theta}\big]^{1}_{1},\, \\ \hspace{0.5cm}p_t^{\text{tot}}=p_t-\alpha \big[T^{\Theta}\big]^{2}_{2}.
\end{eqnarray}

Therefore, total anisotropy is the combination of two different forms of anisotropy that could be expressed as:
\begin{eqnarray} \Delta^{\text{tot}}=p_t^{\text{tot}}-p_r^{\text{tot}}=(p_t-p_r)+\alpha(\Theta^{1}_{1}-\Theta^{2}_{2})\nonumber\\
=\Delta_{f(\mathcal{T})}+\alpha \Delta_{T^{\Theta}}.~~~~
\end{eqnarray}
It is significant to mention that within the present anisotropic compact stellar system, two distinct types of anisotropies are formulated: \(T_{\mu\nu}\) and \(\Theta_{\mu\nu}\). Where, the additional form of anisotropy, \(\Delta_{T^{\Theta}}\), becomes pertinent due to gravitational decoupling, influencing the transformation processes uniquely.
Now, let us focus on the interior of a spherically symmetric static fluid distribution, where the space-time is specified by the following line element:
\begin{eqnarray}\label{metric1}
    dS_{-}^2 = - e^{\nu(r)} dt^2 + e^{\lambda(r)} dr^2
+r^2(d\theta^2+sin^2\theta d\phi ^2).~~
\end{eqnarray}
The above line element results in the distance formula \(dS^2 = g_{ij} dx^{i} dx^{j}\), where \(x^{j} = (t, r, \theta, \phi)\) represents the coordinates of 4D space-time. The functions \(\nu(r)\) and \(\lambda(r)\) denote the static metric potentials along the temporal and radial coordinates, respectively. The torsion scalar and its derivative, expressed in terms of the radial coordinate \(r\), can be calculated as follows:
\begin{eqnarray}\label{torsion}
&&\hspace{-7cm} \mathcal{T}(r)=\frac{2 e^{-\lambda}}{r}\left[\nu^{\prime}+\frac{1}{r}\right], \\  \mathcal{T}^{\prime}(r)=\frac{e^{-\lambda}}{r}\left[\nu^{\prime \prime}-\frac{1}{r^2}-(\nu^{\prime}+\frac{1}{r})(\lambda^{\prime}+\frac{1}{r})\right].
\end{eqnarray}
For the static spherically symmetric metric (\ref{metric1}), the tetrad matrix can be represented as : 
\begin{eqnarray}\label{tetrad}
    e^n_{~~i} =\Big(e^{\frac{\nu}{2}},e^{\frac{\lambda}{2}},r,rsin\theta\Big).
\end{eqnarray}
By substituting the above-mentioned tetrad field (\ref{tetrad}) into Eq.(\ref{field}), the independent component of field equation for an anisotropic fluid in \(f(\mathcal{T})\) gravity can be explicitly derived as follows:
\begin{eqnarray}\label{fe1}
&&\hspace{0cm}8 \pi \rho^{\text{tot}} = \frac{f}{2}+f_\mathcal{T}\bigg[\frac{1}{r^2}+\frac{e^{-\lambda (r)}}{r} \big\{\lambda '(r)+\nu '(r)\big\}-\mathcal{T}(r)\bigg],~~~~~~~\\ 
\label{fe2}
 &&\hspace{0cm}8 \pi p_r^{\text{tot}} = -\frac{f}{2}+f_\mathcal{T}\Big[\mathcal{T}(r)-\frac{1}{r^2}\Big], \\ \label{fe3}
 &&\hspace{0cm}8 \pi p_t^{\text{tot}} = -\frac{f}{2}+f_\mathcal{T} \Bigg[\frac{\mathcal{T}(r)}{2}+e^{-\lambda (r)} \bigg\{\frac{\nu''(r)}{2}+\Big(\frac{\nu '(r)}{4}\nonumber\\&&\hspace{1.5cm}+\frac{1}{2 r}\Big)\big(\nu'(r)-\lambda'(r)\big)\bigg\}\Bigg]. 
\end{eqnarray}
It can be noticed that the above field equation (\ref{fe1}-\ref{fe3}) directly gives the equivalent field equations in GR for the particular form of  $f(\mathcal{T})=\mathcal{T}$. However, in the framework of teleparallel gravity, an extra non-diagonal term comes as follows:
\begin{eqnarray}\label{fe4}
    \frac{\cot\theta}{2 r^2} \mathcal{T^{\prime}}f_{\mathcal{T}\mathcal{T}} = 0 
\end{eqnarray}
There are two possibilities of the above equation (\ref{fe4}) as:\\
\textbf{\text{Case-I}} : $\mathcal{T}^{\prime}=0$ $\Rightarrow \mathcal{T}=\mathcal{T}_0$= \text{constant} that imply $\mathcal{T}$ is free of the radial co-ordinate $r$ therefore, $f(\mathcal{T})$ remains constant.\\
\textbf{{\text{Case-II}}} : $f_{\mathcal{T}\mathcal{T}}=0\implies$  $f(\mathcal{T})=\zeta_1 \mathcal{T}+\zeta_2$, which is the linear form of $f$ where $\zeta_1$ and $\zeta_2$ are two model parameters.

The aforementioned linear function has been effectively utilized in various contexts within \(f(\mathcal{T})\) gravity. Our current aim is to solve the \(f(\mathcal{T})\) gravity field equations (\ref{fe1}-\ref{fe3}) under the functional form \(f(\mathcal{T}) = \zeta_1 \mathcal{T} + \zeta_2\). To accomplish this, we intend to apply the well-established gravitational decoupling method via the complete geometric deformation approach by implementing a complexity-free anisotropic dark star model.

Now, using the torsion scalar (\ref{torsion}) and the linear form of $f(\mathcal{T})$ in Eqs.(\ref{fe1}-\ref{fe3}) we got the three independent component of field equation as :
\begin{eqnarray}\label{FE1}
&&\hspace{0cm}8 \pi \rho^{\text{tot}} = \frac{e^{-\lambda (r)}}{2r^2} \Big[e^{\lambda (r)} (2 \zeta_1+\zeta_2 r^2)-2 \zeta_1+2 \zeta_1 r \lambda'(r)\Big],~~~~~~~
\\ \label{FE2}
&&\hspace{0cm}8\pi p_r^{\text{tot}} = \frac{e^{-\lambda (r)}}{2r^2} \Big[2 \zeta_1-e^{\lambda (r)} (2 \zeta_1+\zeta_2 r^2)+2 \zeta_1 r \nu '(r)\Big],~~~
\\ \label{FE3}
&&\hspace{0cm}8\pi p_t^{\text{tot}} = \frac{e^{-\lambda (r)}}{4r} \Big[-\zeta_1 (r \nu '(r)+2) (\lambda '(r)-\nu '(r))\nonumber\\&&\hspace{2cm}+2 \zeta_1 r \nu ''(r)-2 \zeta_2 r e^{\lambda (r)}\Big].
\end{eqnarray}

The hydrostatic equilibrium equation can be derived by ensuring the continuity of the total stress-energy tensor. This is achieved by setting the divergence of the effective stress-energy tensor to zero, expressed as $\nabla_{\mu}\big[T_{\nu}^{\mu}\big]^{\text{tot}}=0$:

\begin{eqnarray}\label{tov1}
    &&\hspace{-0.1cm}\implies-\frac{\nu^{\prime}}{2}\big(\rho^{\text{tot}}+p_r^{\text{tot}}\big)-\frac{dp_r^{\text{tot}}}{dr}+\frac{2}{r}\big(p_t^{\text{tot}}-p_r^{\text{tot}}\big)=0,\nonumber\\ \label{tov2}
    &&\hspace{-0.1cm}\implies -\frac{dp_r}{dr}-\frac{\nu^{\prime}}{2}(\rho+p_r)+\frac{2}{r}\big(p_t-p_r\big)+\alpha \frac{d \Theta_1^1}{dr}\nonumber\\&&\hspace{0.9cm}-\frac{\nu^{\prime}}{2}\alpha (\Theta_0^0-\Theta_1^1)-\frac{2}{r}\alpha\big(\Theta_2^2-\Theta_1^1\big)=0.
\end{eqnarray}

It is essential to note that equation (\ref{tov1}) represents the well-known Tolman-Oppenheimer-Volkoff (TOV) equation, which governs the decoupling system as described by \cite{tov}. The TOV equations, derived from gravitational theory, are used to describe the structure of a spherically symmetric object in hydrostatic equilibrium.

It can be noticed that as $\alpha \to 0$, the original TOV equations for $f(\mathcal{T})$ gravity are formally recovered. Therefore, our objective is to apply the geometric deformation under the complete gravitational Decoupling (CGD) method to construct the compact star model, aiming to solve the system of equations (\ref{FE1}-\ref{FE3}). In this context, the mass function $\mathcal{M}(r)$ can be expressed as follows:

\begin{eqnarray}\label{mf}
    &&\mathcal{M}(r)=\underbrace{ 4\pi \int_0^{r} \rho(\tilde{r}) \Tilde{r}^2 d\Tilde{r}}_{\mathcal{M}_{\mathcal{T}}(r)}+\underbrace{4\pi \alpha \int_0^{r} \Theta_0^{0}(\Tilde{r}) \Tilde{r}^2\,d\Tilde{r}}_{\mathcal{M}_{C G D}(r)} .
\end{eqnarray}

Apart from that, it will be intriguing to examine how the seed $T_{\mu\nu}$ is physically influenced by the newly generated source $[T_{\theta}]^{i}_j$. This can be achieved through the extended gravitational decoupling approach \cite{gd2}, which requires the necessary transformation of the metric functions (i.e., $e^{\nu}$ and $e^{\lambda}$) as follows:
\begin{eqnarray}\label{mgd1}
&&\hspace{1.3cm}\nu(r) \longrightarrow G(r)+\alpha\, \phi(r),\\ \label{mgd2}
   &&\hspace{1.3cm} e^{-\lambda(r)} \longrightarrow H(r)+\alpha \, \psi (r). 
\end{eqnarray}
Here, $\alpha$ is the decoupling constant. This parameter plays a pivotal role in our study to see the DM influence in the anisotropic strange star. In this particular case, the manipulation of the geometry involves decoupling certain functions in the $t-t$ and $r-r$ components. Specifically, we denote the geometric deformation along the $r-r$ component as $\psi(r)$ and the deformation along the $t-t$ component as $\phi(r)$. In this scenario, we are interested in modifying the $r-r$ component as well as the $t-t$ component. Consequently, we set $\phi(r)\neq0$ and $\psi(r)\neq 0$. This choice highlights the primary objective of this deformation technique. This method is known as \textit{complete (extended) geometric deformation}. Therefore, by implementing the above deformed metric components from equations (\ref{mgd1}) and (\ref{mgd2}) into the field equations (\ref{FE1}-\ref{FE3}), we get two split systems of equations:

\begin{itemize}
    \item The first one is the seed system corresponds to a perfect fluid in teleparallel gravity with $\alpha=0$, having components: $\{\rho, p_r, p_t, G(r), H(r)\}$.
 \begin{eqnarray}\label{sfe1}
&&\hspace{-0.6cm}\rho(r) = \frac{1}{16 \pi r^2}\Big[ 2 \zeta_1-2 \zeta_1 (r H'(r)+H(r))+\zeta_2 r^2\Big],~~
\\ \label{sfe2}
&&\hspace{-0.6cm}p_r = \frac{1}{16\pi r^2}\Big[-2 \zeta_1 H(r) (r G'(r)+1)+2 \zeta_1+\zeta_2 r^2\Big],~~~
\\ \label{sfe3}
&&\hspace{-0.6cm}p_t = \frac{1}{32\pi r}\Big[2 \zeta_1 r H(r) G''(r)+\zeta_1 \big(r G'(r)+2\big)\nonumber\\&& \hspace{1cm}\big(H(r) G'(r)+H'(r)\big)-2 \zeta_2 r\Big].
\end{eqnarray}
\end{itemize}
In these conditions, assuming no exchange of energy-momentum occurs between the perfect fluid $(T_{\mu\nu})$ and the source $\Theta_{\mu\nu}$, and their interaction is purely gravitational, Eq.(\ref{tov2}) yields for the seed system as :
\begin{eqnarray}\label{con1}
    -\frac{dp_r}{dr}-\frac{\mathcal{A^{\prime}}}{2}(\rho+p_r)+\frac{2}{r}\big(p_t-p_r\big)=0,
\end{eqnarray}
Moreover, in this reference, the interior space-time metric can be re-defined as :
\begin{eqnarray}
    dS_{-}^2=-e^{G(r)}dt^2+H^{-1}(r)dr^2+r^2(d\theta^2+sin^2(\theta)d\phi^{2}).~~~~~~
\end{eqnarray}
\begin{itemize}
\item The second set of equations corresponding to the source $\Theta_{\mu\nu}$, possesses components $\{\Theta_0^0,~ \Theta_1^1,~ \Theta_2^2,~ G(r), H(r), \phi(r), \psi(r)\}$, is given by:
\begin{eqnarray}\label{t1}
&&\hspace{0.0cm}\Theta^0_0 = -\frac{\zeta_1}{8\pi r^2}\Big[ \big(r \psi '(r)+\psi (r)\big)\Big],
\\ \label{t2}
&&\hspace{0.0cm}\Theta_1^1 = \frac{-\zeta_1}{8\pi r^2}\Big[ \psi (r) (r G'(r)+1)+r \phi '(r) (H(r)+\alpha  \psi (r))\Big],~~~~~~
\\ \label{t3}
&&\hspace{0.0cm}\Theta^2_2 = \frac{-\zeta_1}{32\pi r\alpha}\Big[-\big(\left(r G'(r)+2\right) (H(r) G'(r)+H'(r))\big)\nonumber\\&&\hspace{0.5cm}+2 r (H(r)+\alpha  \psi (r)) \left(G''(r)+\alpha  \phi ''(r)\right)-2 r H(r)\nonumber\\&&\hspace{0.5cm} G''(r)+\left(r G'(r)+\alpha  r \phi '(r)+2\right) \big((H(r)+\alpha  \psi (r))\nonumber\\&&\hspace{0.5cm} \left(G'(r)+\alpha  \phi '(r)\right)+H'(r)+\alpha  \psi '(r)\big)\Big].
\end{eqnarray}
\end{itemize}

The corresponding continuity equation is given from Eq.(\ref{tov2}) as:
\begin{eqnarray}\label{con2}
     \frac{d \Theta_1^1}{dr}-\frac{\mathcal{A^{\prime}}}{2} (\Theta_0^0-\Theta_1^1)-\frac{2}{r}\big(\Theta_2^2-\Theta_1^1\big)=0.
\end{eqnarray}
These equations(\ref{con1},\ref{con2}) are referred to as the modified TOV equations for the teleparallel gravity and the $\Theta$ gravitational sector, coming from $\nabla_{\mu}T^{\mu\nu}=0$ and $\nabla_{\mu}\Theta^{\mu\nu}=0$, respectively. There are several noteworthy aspects corresponding to the system (\ref{t1}-\ref{t3}). The primary resemblance between this system and the standard spherically symmetric field equations in teleparallel gravity for an anisotropic system with an energy-momentum tensor $\Theta_{\mu\nu}$, where $\{\rho=\Theta_0^0, p_r=\Theta_1^1, p_t=\Theta_2^2\}$ and its conservation equation, is that they are fundamentally alike. Additionally, the active gravitational mass function for both systems can be expressed as:
\begin{eqnarray}
    \mathcal{M}_{\mathcal{T}}(r)=\int_0^r 4 \pi \tilde{r}^2 \rho(\tilde{r})\, d\tilde{r} ;\quad  \mathcal{M}_{\Theta}(r)= \int_0^r 4 \pi \tilde{r}^2 \, \Theta_0^0(\tilde{r}) \, d \tilde{r}\nonumber.
\end{eqnarray}
 Within the framework of teleparallel gravity, the relevant mass functions for the sources $T_{\mu\nu}$ and $\Theta_{\mu\nu}$ are represented as $\mathcal{M}_{\mathcal{T}}(r)$ and $\mathcal{M}_{\Theta}(r)$, respectively. Therefore, in the framework of completely deformed spacetime, the interior mass function can be expressed as:
 \begin{eqnarray}
     \mathcal{M}(r) = \mathcal{M}_{\mathcal{T}}(r) -\frac{\zeta_1 \alpha }{2} r\, \psi(r).
 \end{eqnarray}

\section{COMPLEXITY-FREE ANISOTROPIC DARK STAR MODEL BY GRAVITATIONAL DECOUPLING}\label{seciii}

In this section, let us move to address both sets of Equations (\ref{sfe1}-\ref{sfe3}) and (\ref{t1}-\ref{t3}) corresponding to the sectors $T_{\mu\nu}$ and $\Theta_{\mu\nu}$. The energy-momentum tensor $T_{\mu\nu}$ describes an anisotropic distribution of fluid matter, thereby, $\Theta_{\mu\nu}$ can increase the overall anisotropy within the system, which serves to counteract gravitational collapse. Furthermore, upon scrutiny of the second set of equations, it becomes apparent that the solution to the $\Theta$ sector is contingent upon the resolution of the initial set of equations. Consequently, it is more conventional to solve the initial system first.

From the field equations (\ref{sfe1}-\ref{sfe3}), it can be seen that it is a second-order highly non-linear system of equations, comprising three equations with five unknowns $\{\rho, p_r, p_t, G(r), H(r)\}$. one may independently prescribe any two of these unknowns to ascertain exact solutions. An approach to deriving exact solutions entails fixing one metric potential and imposing an additional assumption (such as a specific equation of state or an embedding condition) to determine another metric potential. However, in this study, we consider the most popular perfect fluid solution given by Tolman-Kuchowicz ansatz \cite{tolman,kucho},
\begin{eqnarray}
   &&\hspace{0.0cm} dS^2=-e^{\text{C} r^2+2 \log \text{D}} dt^2+ (1+\text{A} r^2+\text{B} r^4)^{-1}dr^2\nonumber\\&&\hspace{1.0cm}+r^2(d\theta^2+sin^2{\theta})d\phi^2.
\end{eqnarray}
where $D$ is a dimensionless parameter, and $A$, $B$, and $C$ are constant parameters with units of km$^{-2}$, km$^{-4}$, and km$^{-2}$, respectively. The selection of metric potentials in this study is well-justified as they give a model devoid of singularities. This potential has been utilized to explore many stellar models in the literature \cite{sp3,pb3}.

Now, for the above space-time solution, we get the form of the seed energy density, radial pressure, and tangential pressure given by:
\begin{eqnarray}
    &&\hspace{0.0cm}\rho= -3 \text{A} \zeta_1-5 \text{B} \zeta_1 r^2+\frac{\zeta_2}{2},\\
    &&\hspace{0.0cm}p_r=\zeta_1 \left(r^2 (2 \text{A} \text{C}+\text{B})+\text{A}+2 \text{B} \text{C} r^4+2 \text{C}\right)-\frac{\zeta_2}{2},\\
    &&\hspace{0.0cm}p_t=\zeta_1 \big(\text{C} r^4 (\text{A} \text{C}+4 \text{B})+r^2 (\text{C} (3 \text{A}+\text{C})+2 \text{B})+\text{A}\nonumber\\&&\hspace{1.0cm}+\text{B} \text{C}^2 r^6+2 \text{C}\big)-\frac{\zeta_2}{2}.
\end{eqnarray}
Next, we proceed to determine the solution of the $\Theta$ sector, which constitutes the principal strength of this manuscript. From Equations (\ref{t1}-\ref{t3}), it is apparent that, after deriving the seed solution, the system now encompasses the unknown parameters $\{\Theta_0^0,~ \Theta_1^1,~ \Theta_2^2,~ \phi(r), \psi(r)\}$. To close the system of differential equations and determine $\phi$ and $\psi$, it is imperative to solve for these parameters. Given our interest in a dark star with zero complexity factor, it is essential to first define the condition for vanishing complexity in a gravitationally decoupled system. We use Herrera's definition of complexity \cite{gd10} to develop the vanishing complexity requirement in the context of teleparallel-gravity. His definition provides the following definition for the complexity factor $(Z^{\mathcal{T}}_{TF})$ for the system of equation (\ref{fe1}-\ref{fe3}).
\begin{eqnarray}
    Z_{TF}^{\mathcal{T}}=8\pi(p_r^{\text{tot}}-p_t^{\text{tot}})-\frac{4\pi}{2r^3}\int^r_0 y^3 (\rho^{\text{tot}})^{\prime}(y)dy.
\end{eqnarray}
Now by implementing the form of $\rho_r^{tot}, p_r^{tot} $ and $p_t^{tot}$
in the above factor, it gives,
\begin{eqnarray}
     &&\hspace{0cm}Z_{TF}^{\mathcal{T}}=\frac{\zeta_1}{4r e^{\lambda}}\big[\nu^{\prime}\{r(\lambda^{\prime}-\nu^{\prime})+2\}-2r\nu^{\prime \prime}\big]
\end{eqnarray}

Now, vanishing the complexity factor gives two possibilities. Since $\zeta_1$ can not be zero, Therefore,
\begin{eqnarray}
     &&\hspace{0cm}Z_{TF}^{\mathcal{T}}=0\\
    &&\hspace{-0.9cm} \implies \nu^{\prime}\{r(\lambda^{\prime}-\nu^{\prime})+2\}-2r\nu^{\prime \prime}=0
\end{eqnarray}
After performing the integration, the first-order solution of the above differential equation can
be written as :
\begin{eqnarray}
    \nu^{\prime} e^{\nu/2}= \mathcal{C}_1 r e^{\lambda/2}
\end{eqnarray}

Finally, the above equation provides the bridge between two metric potentials $\nu$ and $\lambda$ as
\begin{eqnarray}
    &&\hspace{0.0cm}\nu(r)= 2\text{Log}[\mathcal{C}_2+\mathcal{C}_1\int  r e^{\lambda/2} dr],\\
    \label{phie}&&\hspace{-0.2cm}\implies\nu(r)=2\text{Log}\big[\mathcal{C}_2+\mathcal{C}_1\int  \frac{r}{\sqrt{H(r)+\alpha\psi(r)}}  dr\Big]
\end{eqnarray}

Where, $\mathcal{C}_1$ and $\mathcal{C}_2$ are integrating constants to be determined by the matching condition. Interestingly, the relation found in teleparallel gravity theory is comparable to the one found in the setting of Einstein's general relativity \cite{cf1} by Contreras and Stuchlik.

we must now determine the deformation function $\phi(r)$ and $\psi(r)$ to get the solution of $\Theta_0^0$, $\Theta_1^1$ and $\Theta_2^2$. As previously mentioned, the primary focus of this study is the vanishing complexity approach to constructing the dark star model. NSs, the densest objects composed of unknown particles, may offer insights into unexpected varieties of DM through unforeseen interactions. Their extreme properties have been leveraged in diverse DM searches. Recent studies of high-resolution optical rotation curves of galaxies focus on uncovering precise model parameters for galactic discs embedded within cold dark matter (CDM) halos. By comparing Navarro-Frenk-White (NFW) and pseudo-isothermal profiles, researchers found that $68\%$ of the galaxies are best described by a pseudo-isothermal profile with a core, suggesting it provides a superior fit for these systems \cite{dm1}. In our analysis for fitting the dark-star models, we use the pseudo-isothermal dark matter energy density distribution to mimic $\Theta_0^0$ given by,
\begin{eqnarray}
    \rho_{DM}=\rho_0\big[1+\big(\frac{r}{r_0}\big)^2\big]^{-1}.
\end{eqnarray}
Additionally, the study results for $\rho_0$ and $r_0$ as halo density and halo core radius may be found in reference \cite{dm1},\cite{dm2} on dark matter haloes and rotation curves of galaxies. However, in the present investigation, we treated the constants $L=\rho_0$ and $N=1/r_0^2$ as free parameters. It's noticeable that the density profile $\rho_{DM}$ is both regular and free from singularities. Additionally, it decreases monotonically as the radial coordinate $r$ increases inside any finite region. So, for this dark matter profile, the deformation function $\psi(r)$ can be obtained from the equation \eqref{t1} after the integration as:
\begin{eqnarray}
    \psi(r)=-\frac{L}{r} \left(\frac{r}{\zeta_1 N}-\frac{\tan ^{-1}\left(\sqrt{N} r\right)}{\zeta_1 N^{3/2}}\right)+\frac{\mathcal{C}_3}{r}.
\end{eqnarray}
The aforementioned solution naturally yields $\mathcal{C}_3=0$ as we are interested in a non-singular system. Again, we may expand in a series around to $r = 0$ to remove the singular characteristic because the $tan^{-1}\left(\sqrt{N} r\right)$ is continuous within the range $[0, R]$. Thus, by taking into account up to $\mathcal{O}({r^5})$, we are ultimately able to determine the result for $\psi(r)$ as :
\begin{eqnarray}
   \psi(r)= \frac{L r^2 \left(3 N r^2-5\right)}{15 \zeta_1}.
\end{eqnarray}
Furthermore, we can determine the deformation function $\phi(r)$ for the given seed metric functions$ ( { G(r) , H(r) } ) $ and the radial deformation function $\psi(r)$ by utilizing the equation (\ref{phie}) as follows:
\begin{eqnarray}
    &&\hspace{0.0cm}\phi(r)=\frac{1}{\alpha}\Big\{2\text{Log}[\mathcal{C}_2+\mathcal{C}_1\int  \frac{r}{\sqrt{H(r)+\alpha\psi(r)}}  dr]-G(r)\Big\}\nonumber\\ \label{phi}
    &&\hspace{-0.0cm} \implies \phi(r)= \frac{1}{\alpha}\bigg[-C r^2-2 \log (D)+2\text{Log}\Big[\mathcal{C}_2-\nonumber\\&&\hspace{0.9cm}\frac{\mathcal{C}_1 \zeta_1}{\sqrt{5} (\alpha  L-3 A \zeta_1)}\Big(\big\{\big(45 \left(A r^2+B r^4+1\right)+3 \alpha \zeta_1^{-1}\nonumber\\&&\hspace{0.4cm} L r^2 \left(3 N r^2-5\right)\big)^{1/2}-3 r^2 \sqrt{5 B+\frac{\alpha  L N}{\zeta_1}}\big\}\Big)-\frac{1}{4}\nonumber\\&&\hspace{0.4cm}\big(B+\frac{\alpha  L N}{5 \zeta_1}\big)^{1/2}\Big(\mathcal{C}_1\text{Log}\big[-225 A^2 \zeta_1^2+900 A B \zeta_1^2 r^2\nonumber\\&&\hspace{0.4cm}+150 \alpha  A \zeta_1 L+180 \alpha  A \zeta_1 L N r^2+1800 B^2 \zeta_1^2 r^4+900 B \zeta_1^2\nonumber\\&&\hspace{0.4cm}+720 \alpha  B \zeta_1 L N r^4-300 \alpha  B \zeta_1 L r^2-25 \alpha ^2 L^2+72 \alpha ^2 L^2 N^2 r^4\nonumber\\&&\hspace{0.4cm}-60 \alpha ^2 L^2 N r^2+180 \alpha  \zeta_1 L N-24 r^2\zeta_1 \sqrt{15 B+\frac{3 \alpha  L N}{\zeta_1}}(5 B\zeta_1\nonumber\\&&\hspace{0.4cm}+\alpha  L N) \sqrt{15 \left(A r^2+B r^4+1\right)+\frac{\alpha  L r^2 \left(3 N r^2-5\right)}{\zeta_1}}\big]\Big)\Big]\bigg]~~~~~~~
\end{eqnarray}

By using the above deformation function $\phi(r)$ and $\psi(r)$, the remaining components of $\Theta$-sector can be obtained from Eqs. (\ref{t1},\ref{t2},\ref{t3}).

Therefore, for the above solution, the final form of metric potentials is,

\begin{eqnarray}
   &&\hspace{0.0cm} e^{-\lambda}(r)=(1+\text{A} r^2+\text{B} r^4)+\alpha\frac{  \big\{\text{L} r^2 (3 \text{N} r^2-5)\big\}}{15\zeta_1}.\\
   &&\hspace{0.0cm} \nu(r)= 2\text{Log}\Big[\mathcal{C}_2-\frac{\mathcal{C}_1 \zeta_1}{\sqrt{5} (\alpha  L-3 A \zeta_1)}\Big(\big\{\big(45 (A r^2+B r^4\nonumber\\&&\hspace{0.9cm}+1)+3 \alpha \zeta_1^{-1} L r^2 \left(3 N r^2-5\right)\big)^{1/2}-3 r^2 (5 B+\nonumber\\&&\hspace{0.9cm}\frac{\alpha  L N}{\zeta_1}\big)^{1/2}\big\}\Big)-\frac{1}{4}\big(B+\frac{\alpha  L N}{5 \zeta_1}\big)^{1/2}\Big(\mathcal{C}_1\text{Log}\big[-225\nonumber\\&&\hspace{0.9cm} A^2 \zeta_1^2+900 A B \zeta_1^2 r^2+150 \alpha  A \zeta_1 L+180 \alpha  A \zeta_1 L N r^2\nonumber\\&&\hspace{0.9cm}+1800 B^2 \zeta_1^2 r^4+900 B \zeta_1^2+720 \alpha  B \zeta_1 L N r^4-300 \alpha  B\nonumber\\&&\hspace{0.9cm} \zeta_1 L r^2-25 \alpha ^2 L^2+72 \alpha ^2 L^2 N^2 r^4-60 \alpha ^2 L^2 N r^2+180\nonumber\\&&\hspace{0.9cm} \alpha  \zeta_1 L N-24 r^2\zeta_1 \sqrt{15 B+\frac{3 \alpha  L N}{\zeta_1}}(5 B\zeta_1+\alpha  L N)\nonumber\\&&\hspace{0.9cm} \sqrt{15 \left(A r^2+B r^4+1\right)+\frac{\alpha  L r^2 \left(3 N r^2-5\right)}{\zeta_1}}\big]\Big)\Big].
\end{eqnarray}
Furthermore, the final expression of energy density, radial pressure, and transverse pressure is given by,
\begin{eqnarray}
    &&\hspace{0.0cm}\rho^{tot}=-3 A \zeta_1-5 B \zeta_1 r^2+\frac{\zeta_2}{2}+\frac{\alpha  L}{N r^2+1},\label{54}\\
   &&\hspace{0.0cm} p_r^{tot}=\zeta_1 \Big(A \left(2 C r^2+\alpha  \phi_1(r) r+1\right)+\frac{1}{r}\left(B r^4+1\right) (2 C r\nonumber\\&&\hspace{0.9cm} +\alpha  \phi_1(r))+B r^2\Big)+\frac{1}{15} \alpha  L \left(3 N r^2-5\right) \big(2 C r^2+\alpha \nonumber\\&&\hspace{0.9cm}  \phi_1(r) r+1\big)-\frac{\zeta_2}{2},\label{55}\\
   &&\hspace{0.0cm} p_t^{tot}=\frac{1}{60} \alpha  L \Big(r \big(4 C^2 r^3 \left(3 N r^2-5\right)+4 C r (3 N r^2 (\alpha  \phi_1(r) r\nonumber\\&&\hspace{0.9cm} +4)-5 (\alpha  \phi_1(r) r+3))+3 N r (\alpha  r (\phi_1(r) (\alpha  \phi_1(r) r+6)\nonumber\\&&\hspace{0.9cm} +2 \phi_2(r) r)+8)-5 \alpha  (\phi_1(r) (\alpha  \phi_1(r) r +4)+2 \phi_2(r) r)\Big)\nonumber\\&&\hspace{0.9cm}-20\big)+A \zeta_1-\frac{\zeta_2}{2}+\frac{1}{4r}\Big\{\zeta_1 \Big(4 C^2 \left(A r^5+B r^7+r^3\right)\nonumber\\&&\hspace{0.9cm}+4 C r \big(r \big(A r (\alpha  \phi_1(r) r+3)+B r^3 (\alpha  \phi_1(r) r+4)+\alpha  \nonumber\\&&\hspace{0.9cm}\phi_1(r)\big)+2\big)+\alpha  r \big(\phi_1(r) (A r (\alpha  \phi_1(r) r+4)+\alpha  \phi_1(r))\nonumber\\&&\hspace{0.9cm}+2 \phi_2(r) \left(A r^2+1\right)\big)+B r^3 (\alpha  r (\phi_1(r) (\alpha  \phi_1(r) r+6)\nonumber\\&&\hspace{0.9cm}+2 \phi_2(r) r)+8)+2 \alpha  \phi_1(r)\Big)\Big\}.\label{56}
\end{eqnarray}
Where $\phi_1(r)$ and $\phi_2(r)$ is given by $\frac{d\phi}{dr}$ and $\frac{d^2\phi}{dr^2}$ respectively. 

\subsection{Matching Condition}

Any stable celestial object with a spherically symmetric structure must have a smooth and continuous stellar distribution at its surface $r=r_{\Sigma}$, extending between its interior $r<r_{\Sigma}$ and exterior solutions $r>r_{\Sigma}$. It can be provided by matching the exterior and interior metrics, or at $r=r_{\Sigma}$. for matching with the outer region of space-time, we have considered the well-known Schwarzschild metric :
\begin{eqnarray}\label{+}
    dS_{+}^2 = -\Big(1-\frac{2\mathcal{M}}{r}\Big) dt^2 + \Big(1-\frac{2\mathcal{M}}{r}\Big)^{-1} dr^2 \nonumber\\&&\hspace{-6.5cm} + ~r^2 (d\theta^2+sin^2 \theta d\phi^2).
\end{eqnarray} Therefore, the boundary condition can be used to find the solution for the unknown constants. So, the continuity of the first fundamental form of boundary condition gives,

\begin{eqnarray}\label{b1}
&&\hspace{0.0cm}g_{tt}(r_{\Sigma})^+=g_{tt}(r_{\Sigma})^-\,\,;\quad g_{rr}(r_{\Sigma})^+=g_{rr}(r_{\Sigma})^-\\
&&\hspace{-0.5cm}\implies(1-\frac{2M}{R_{\Sigma}})=\text{Exp}[C R_{\Sigma}^2+2\text{Log}D+\alpha \phi(R_{\Sigma})]\\ \label{b2}
  &&\hspace{-0.1cm}  1-\frac{2M}{R_{\Sigma}}= (1+\text{A} R_{\Sigma}^2+\text{B} R_{\Sigma}^4)+\alpha\frac{\text{L} R_{\Sigma}^2 (3 \text{N} R_{\Sigma}^2-5)}{15\zeta_1}~~
\end{eqnarray}
Moreover, the continuity of the second fundamental form says, the total radial pressure of a self-gravitating object must be zero at the stellar boundary, i.e. 
\begin{eqnarray}
    &&\hspace{0.0cm}p_r^{tot}\Big|_{r=r_{\Sigma}}=0\\
     &&\hspace{-0.5cm}\implies p_r-\alpha\Theta_1^1\Big|_{r=r_{\Sigma}}=0\\
    &&\hspace{-0.5cm}\implies p_r(r_{\Sigma})+\frac{\zeta_1}{8\pi r_{\Sigma}^2}\Big[ \psi (r_{\Sigma}) (r_{\Sigma} G'(r_{\Sigma})+1)\nonumber\\&&\hspace{1.0cm}+r_{\Sigma} \phi '(r_{\Sigma}) (H(r_{\Sigma})+\alpha  \psi (r_{\Sigma}))\Big]=0.\label{59}
\end{eqnarray}
By solving the above system (\ref{b1}), (\ref{b2}) and (\ref{59}) we can determine the solution for the integrating constants $\mathcal{C}_1$ and $\mathcal{C}_2$.
In the next section, we will discuss the astrophysical implementation of our constructed dark star model via physical analysis and stability analysis.

\section{PHYSICAL ANALYSIS OF THE STELLAR MODEL}\label{seciv}
In this section, we shall examine various physical properties of the stellar dark star model to present the best model through our analysis. Let us start with the important and well-known fundamental energy profiles of the stellar star in the following subsections.

\begin{figure*}
    \centering
    \includegraphics[height=6.2cm,width=7.8cm]{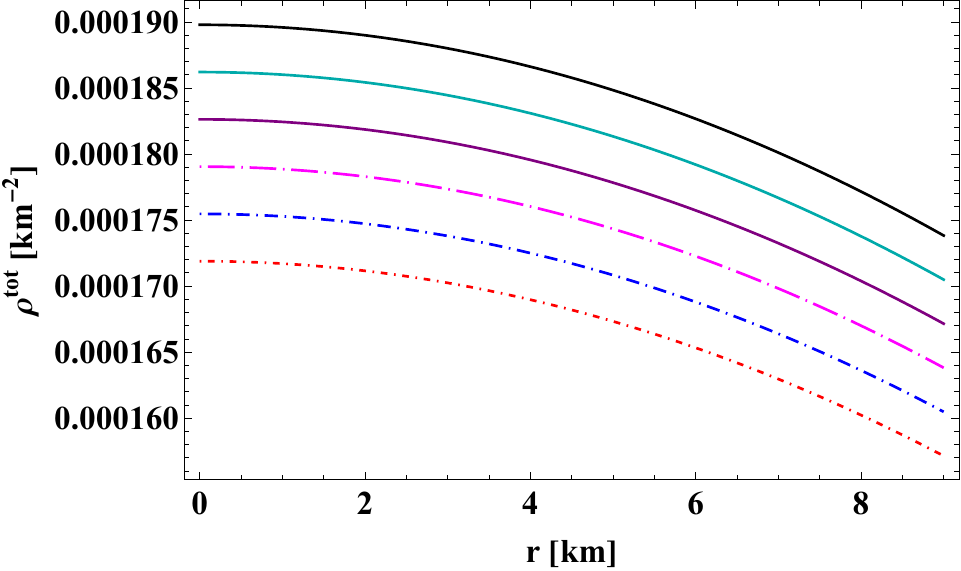}~~~
    \includegraphics[height=6.2cm,width=8cm]{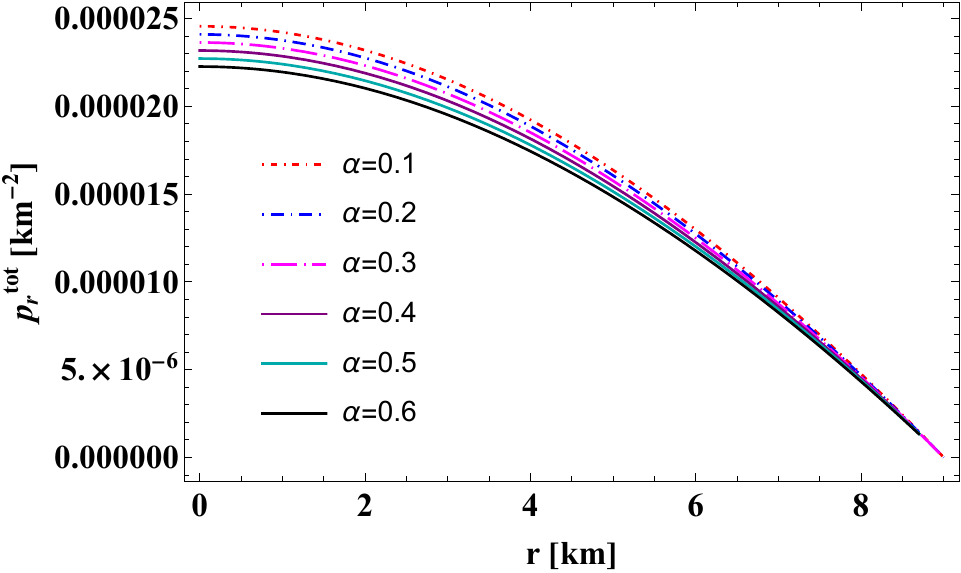}
    \caption{Graphical analysis of energy density (left) [$\alpha=0.1(\textcolor{red}\star),\alpha=0.2(\textcolor{blue}\star), \alpha=0.3(\textcolor{magenta}\star), \alpha=0.4(\textcolor{purple}\star), \alpha=0.5(\textcolor{green}\star),\alpha=0.6(\textcolor{black}\star) $] and radial pressure (right) for $C = 0.288~ \text{km}^{-2}; D = 0.1; A = 0.009~ \text{km}^{-2}; B = 0.000009~\text{km}^{-4}; L = 0.0009 ~\text{km}^{-2}; N = 0.0009~ \text{km}^{-2}$.}
    \label{fig1}
\end{figure*}

\begin{figure*}
    \centering
    \includegraphics[height=6.2cm,width=8cm]{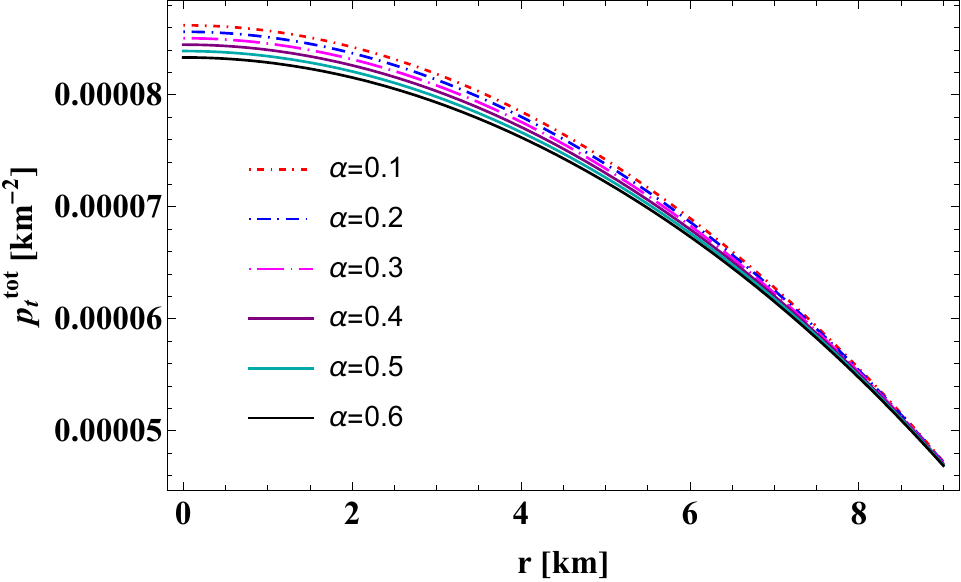}~~~
    \includegraphics[height=6.2cm,width=8cm]{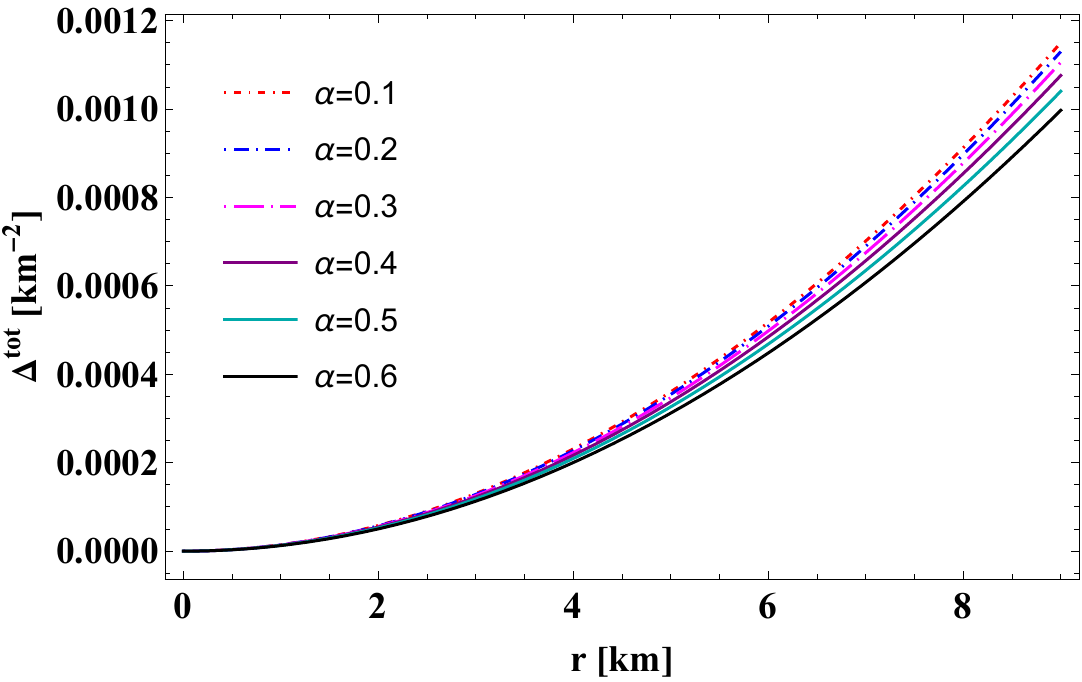}
    \caption{Graphical analysis of tangential pressure (left) and anisotropy (right) for $C = 0.288~ \text{km}^{-2}; D = 0.1; A = 0.009~ \text{km}^{-2}; B = 0.000009~\text{km}^{-4}; L = 0.0009 ~\text{km}^{-2}; N = 0.0009~ \text{km}^{-2}$.}
    \label{fig2}
\end{figure*}

\subsection{Density, pressure, and anisotropy}
This subsection will explore various energy distribution profiles for the above-analyzed star model configured with the perfect fluid and dark matter components. The quantities such as energy density ($\rho^{tot}$), radial pressure ($p_r^{tot}$), and transverse pressure ($p_t^{tot}$) are going to help us to discuss the energy conditions, which represented in Equations \eqref{54}, \eqref{55}, and \eqref{56}, respectively.

The graphical representation of $\rho^{tot}$, $p_r^{tot}$, $p_t^{tot}$, and anisotropy ($\Delta^{tot}$) in Figures \ref{fig1} and \ref{fig2} provides a comprehensive view of the energy distribution within the stellar model. Our analysis reveals that $\rho^{tot}$, $p_r^{tot}$, and $p_t^{tot}$ satisfy the essential criteria for a valid stellar model. Specifically, these quantities are positive and finite throughout the stellar region, peaking at the center and decreasing as the radial distance increases. As depicted in Figure \ref{fig1}, the total energy density ($\rho^{\text{tot}}$) and radial pressure $p_r^{\text{tot}}$ reach zero at the stellar surface ($r = r_{\Sigma}$). Moreover, Figure \ref{fig1} shows that the anisotropy ($\Delta^{tot}$), defined as the difference between the tangential and radial pressures ($p_t^{tot} - p_r^{tot}$), is positive and increases with radial distance. This positive anisotropy, where $p_r^{tot} < p_t^{tot}$, is crucial as it contributes to the stability of the stellar system by supporting hydrostatic equilibrium.

Let's delve more deeply into the behavior of the physical quantities $\rho^{tot}$, $p_r^{tot}$, and $p_t^{tot}$ with the changes of decoupling parameter $\alpha$, as depicted in Figures \ref{fig1}-\ref{fig2}. These figures reveal a fascinating trend: with each increment in $\alpha$, the energy density throughout the anisotropic dark star rises. It is suggested that by utilizing the new source term $\Theta_{\mu\nu}$, our model's dark matter component $\alpha$ leads to the formation of denser stellar objects. Examining Figure \ref{fig1}, we see an opposite behavior with the total energy density in the radial pressure. In the central region, the radial pressure decreases with higher $\alpha$, then it converges near the surface and eventually drops to zero at around 9 km. A similar pattern is observed in Figure \ref{fig2} for the tangential pressure, which converges to different finite, non-zero values at the surface for each $\alpha$. This convergence indicates that the surface pressures, both radial and tangential, remain unaffected by changes in $\alpha$, producing effects akin to those seen in the isotropic systems. Figure \ref{fig2} provides further insights, showing that the anisotropy introduced by the dark matter component intensifies near the surface and converges towards the center as $\alpha$ increases. This demonstrates a monotonically increasing effect of $\alpha$ on the star's anisotropy from the center to the surface. Additionally, the presence of dark matter sets critical limits on the energy density, with $\rho(0) \approx \mathcal{O}( 10^{14}) \, \text{g/cm}^3$ and $\rho(r_{\Sigma})\approx \mathcal{O}( 10^{14}) \, \text{g/cm}^3$ for positive values of $\alpha$. Also, $p_r^{\text{tot}}(0) \approx \mathcal{O}( 10^{14}) \, \text{g/cm}^3$ and $p_t(r_{\Sigma})\approx \mathcal{O}( 10^{14}) \, \text{g/cm}^3$ for positive values of $\alpha$. Here, $\rho(0) = \rho_c$ denotes the central density, while $r_{\Sigma}$ represents the surface of the star.

\subsection{Energy conditions and gradients }\label{seciva}
This subsection will explore whether the anisotropic matter with a dark matter component in the dark stellar configuration satisfies the necessary energy conditions: null energy condition (NEC), weak energy condition (WEC), dominant energy condition (DEC), and strong energy condition (SEC). Specifically:

\begin{itemize}
\item NEC: $\rho^{\text{tot}}+p_r^{\text{tot}} \geq 0,\,\, \rho^{\text{tot}}+p_t^{\text{tot}} \geq 0$,\\
\item WEC: $\rho^{\text{tot}} \geq 0,\,\,\rho^{\text{tot}}+p_r^{\text{tot}} \geq 0,\,\, \rho^{\text{tot}}+p_t^{\text{tot}} \geq 0$,\\
\item DEC: $\rho^{\text{tot}} \geq 0,\,\,\rho^{\text{tot}}-|p_r^{\text{tot}}| \geq 0,\,\, \rho^{\text{tot}}-|p_t^{\text{tot}}| \geq 0$,\\
\item SEC: $\rho^{\text{tot}}+p_r^{\text{tot}}+p_t^{\text{tot}} \geq 0,\,\,\rho^{\text{tot}}+p_r^{\text{tot}} \geq 0,\,\, \rho^{\text{tot}}+p_t^{\text{tot}} \geq 0$.
\end{itemize}

Practically, the NEC, WEC, and DEC confirm that the energy density observed by any observer is always positive, while the SEC implies gravitational attraction. Figures \ref{fig1} and \ref{fig2} show that the NEC, WEC, and SEC are met, as the energy density, radial pressure, and tangential pressure are positive and finite throughout the star. These figures also confirm that the DEC is satisfied. The decoupling constant $\alpha$ significantly influences these energy conditions, making the inequalities more positive as $\alpha$ increases. Thus, as the strength of the dark matter component increases, the energy conditions are more effectively satisfied in the dark star model.

Along with these conditions, we examine the gradients of the energy density, the radial and tangential pressures for our stellar star model. They have to satisfy the following conditions to be a realistic stellar object, defined as
\begin{eqnarray}
    \frac{d\rho^{\text{tot}}}{dr}<0,\,\,\,\frac{dp_r^{\text{tot}}}{dr}<0,\,\,\,\frac{dp_t^{\text{tot}}}{dr}<0.
\end{eqnarray}

\begin{figure*}
    \centering
    \includegraphics[height=4.8cm,width=6.0cm]{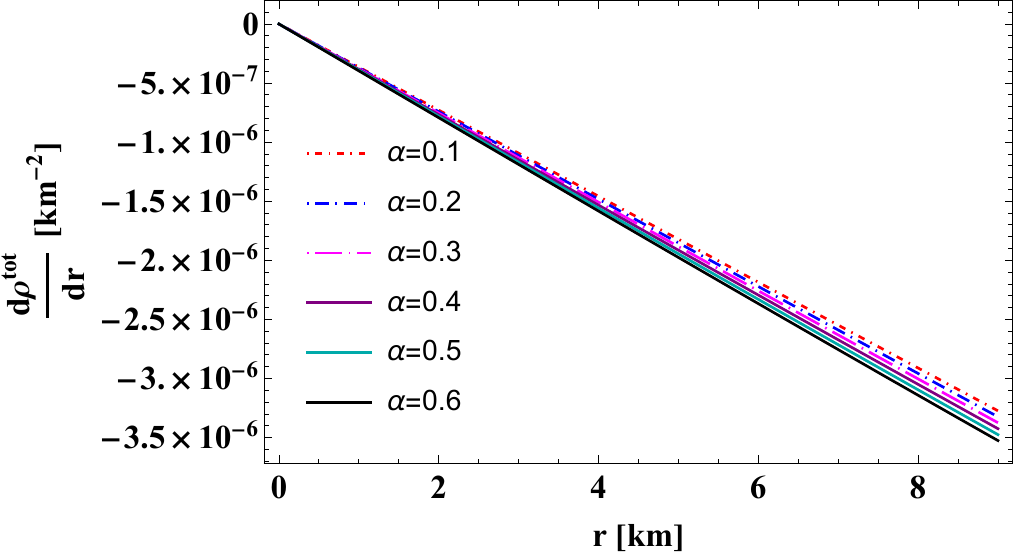}
    \includegraphics[height=4.8cm,width=5.8cm]{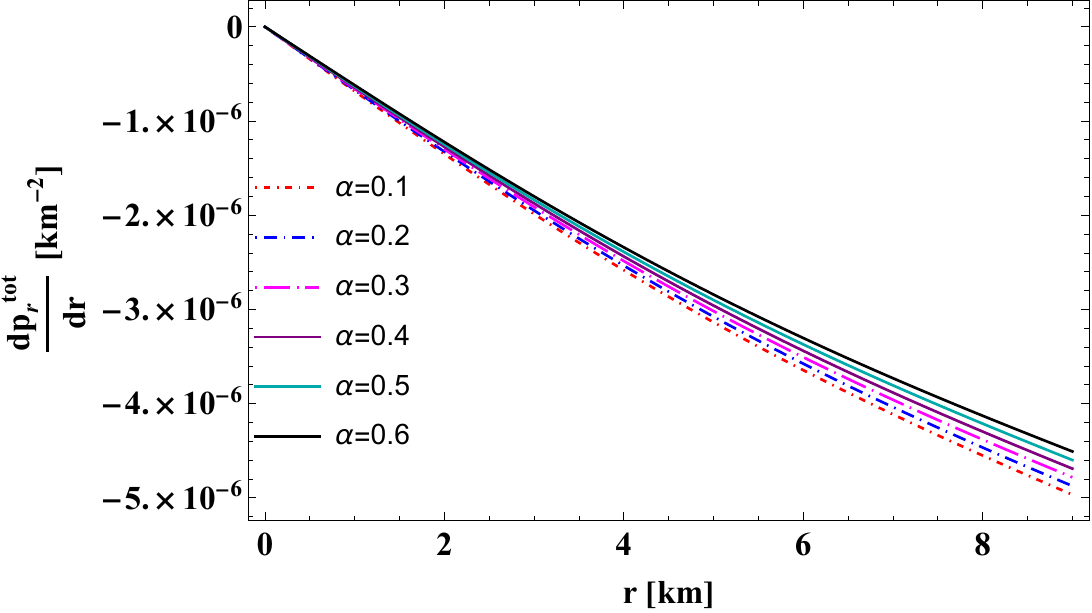}
     \includegraphics[height=4.8cm,width=5.8cm]{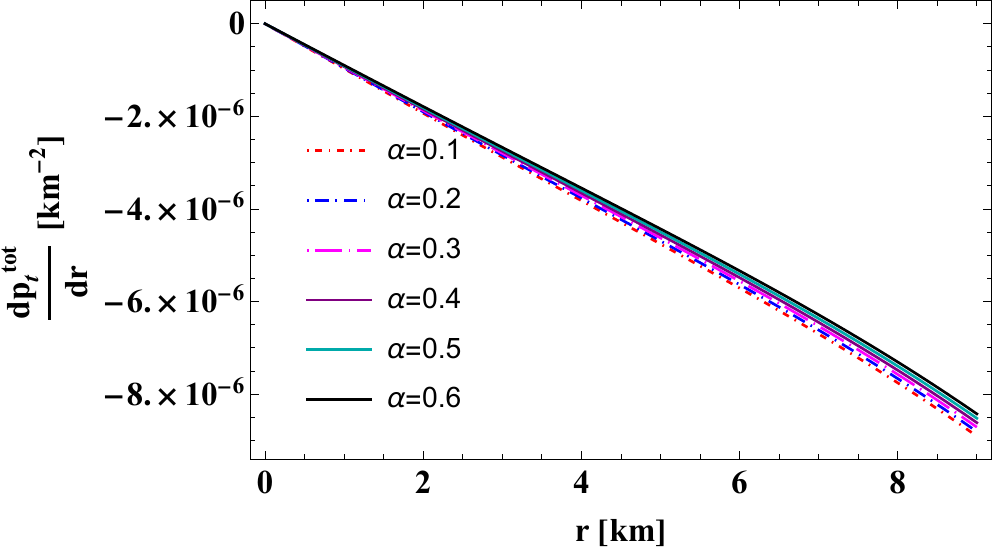}
    \caption{Graphical analysis of density gradient (left), radial pressure gradient (middle) and tangential pressure gradient (right) w.r.t '$r$,' for $C = 0.288~ \text{km}^{-2}; D = 0.1; A = 0.009~ \text{km}^{-2}; B = 0.000009~\text{km}^{-4}; L = 0.0009 ~\text{km}^{-2}; N = 0.0009~ \text{km}^{-2}$.}
    \label{fig4}
\end{figure*}

We graphically present the behavior of the gradients in Figure \ref{fig4}. One can easily observe that the gradients satisfy the above-mentioned conditions and vanish at $r=0$, i.e.,

\begin{eqnarray}
    \frac{d\rho^{\text{tot}}}{dr}=0|_{r=0},\,\,\,\frac{dp_r^{\text{tot}}}{dr}=0|_{r=0},\,\,\,\frac{dp_t^{\text{tot}}}{dr}=0|_{r=0}.
\end{eqnarray}

\section{Stability analysis}\label{secv}

The anisotropic fluid distributions of the dark stellar structure are able to verify the stability of this type of structure; these distributions can be examined through various stability analyses, and some of these, such as Adiabatic index, speed of sound through the dark matter fluid, and analysis of central density, shall be discussed in the following subsections.

\subsection{Stability analysis via Adiabatic Index}\label{secva}

Let us start with analyzing the hydrostatic equilibrium of our stellar dark star configuration by studying the adiabatic index ($\Gamma^{\text{tot}}$). The adiabatic index is formulated as the ratio of two specific heats for a relativistic anisotropic stellar structure and is defined as :
\begin{equation}
    \Gamma^{\text{tot}}= \frac{\rho^{\text{tot}}+p_r^{\text{tot}}}{p_r^{\text{tot}}}\times \frac{dp_r^{\text{tot}}}{d\rho^{\text{tot}}}.
\end{equation}

\begin{figure}
    \centering
    \includegraphics[height=6.2cm,width=7.8cm]{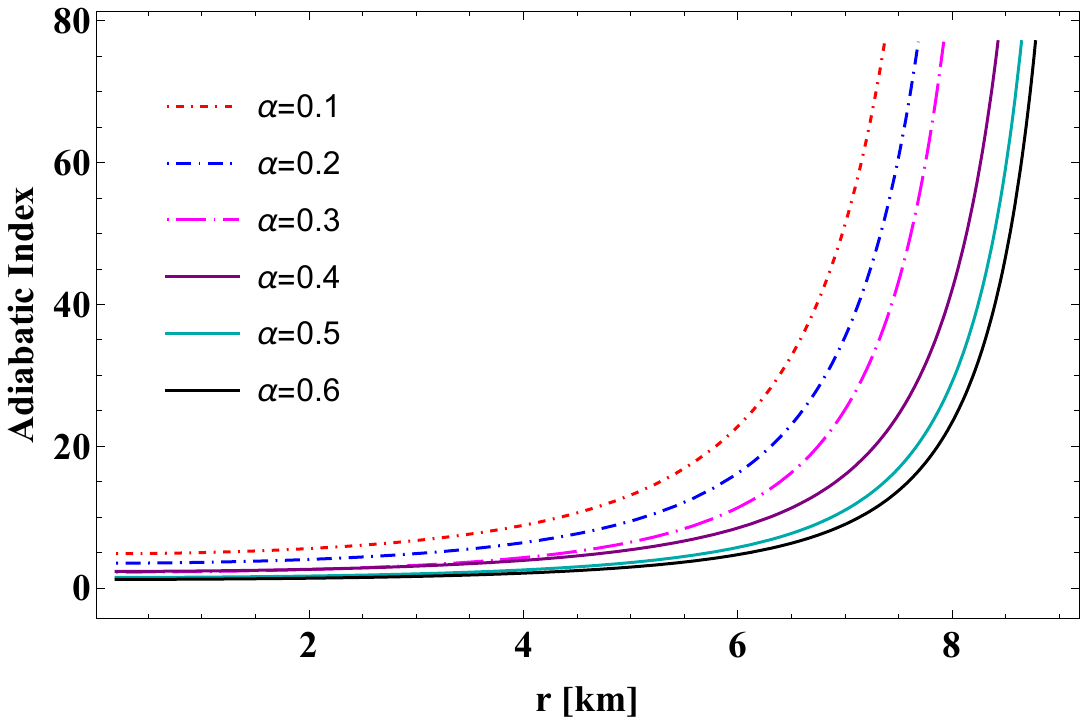}~~~
    \caption{Stability analysis via adiabatic index ($\Gamma$) for $C = 0.288~ \text{km}^{-2}; D = 0.1; A = 0.009~ \text{km}^{-2}; B = 0.000009~\text{km}^{-4}; L = 0.0009 ~\text{km}^{-2}; N = 0.0009~ \text{km}^{-2}$.}
    \label{fig3}
\end{figure}

The stability condition for an isotropic fluid is $\Gamma^{\text{tot}}>4/3$ in the Newtonian limit, whereas for neutral equilibrium, it becomes $\Gamma^{\text{tot}}=4/3$ \cite{bondi/1964}. Later, a study by Heintzmann and Hillebrandth \cite{hei/1975} suggested that the adiabatic index must be $\Gamma^{\text{tot}}>4/3$ for an anisotropic stellar object to be in equilibrium. Also, they analyzed that the positive anisotropy factor helps to improve the limit for $\Gamma^{\text{tot}}$ for an anisotropic dark stellar star in the stability condition. From Figure \ref{fig3}, we verify that the adiabatic index ($\Gamma^{\text{tot}}$) shows the values of greater than $4/3$ for different values of decoupling constant $\alpha$. In addition, we observed an interesting feature from the analysis of the anisotropic factor in Figure \ref{fig2} and adiabatic index in Figure \ref{fig3} that even though the values of decoupling constant $\alpha$ have nearly no effect on $\Delta^{\text{tot}}$ at the neighborhood of the center of the stellar star, whereas it has a significant effect on $\Gamma^{\text{tot}}$ for small radius. Furthermore, it can be noticed that the small values of the decoupling constant $\alpha$ are able to mimic the stability of the anisotropic system near the center. While at the surface region of the star, it has no effect because $\Gamma^{\text{tot}}>4/3$ for all positive values of $\alpha$.

\subsection{Causality Criterion, Herrera's cracking method}\label{secvb}

\begin{figure*}
    \centering
    \includegraphics[height=6.2cm,width=7.8cm]{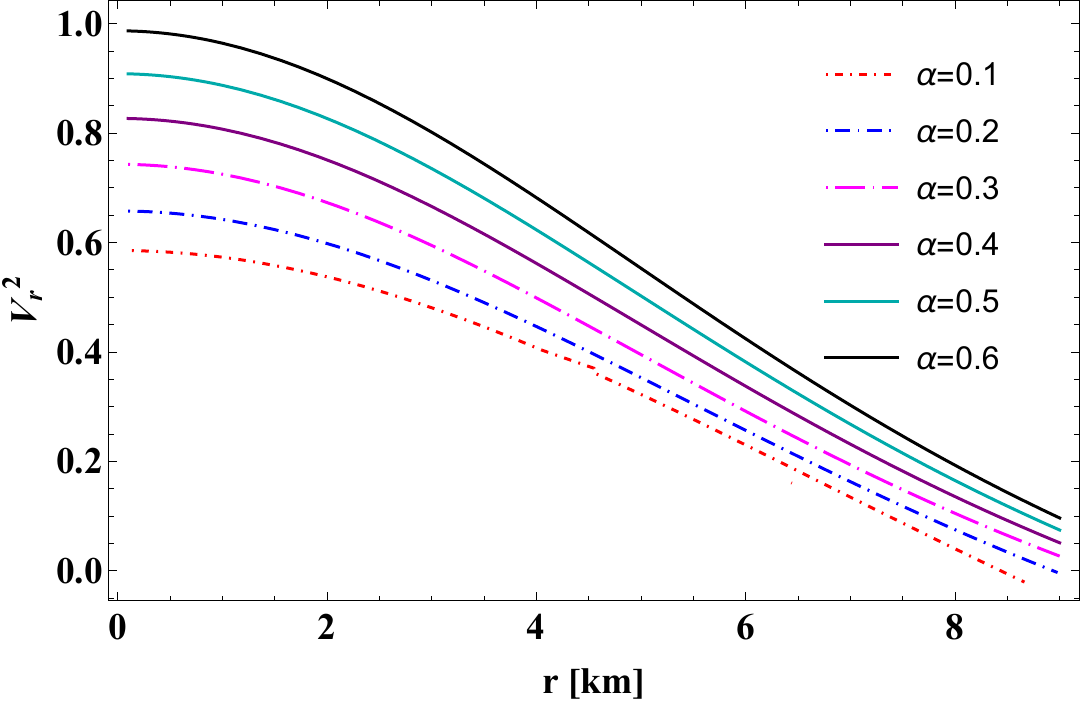}
    \includegraphics[height=6.2cm,width=8cm]{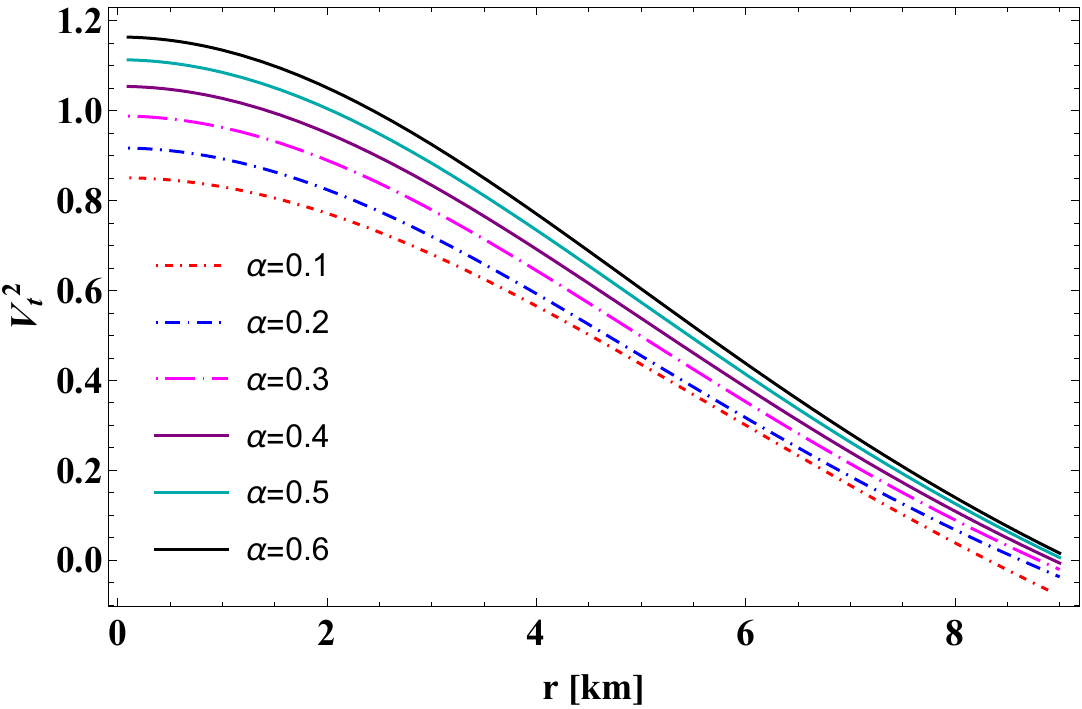}
     \includegraphics[height=6.2cm,width=8cm]{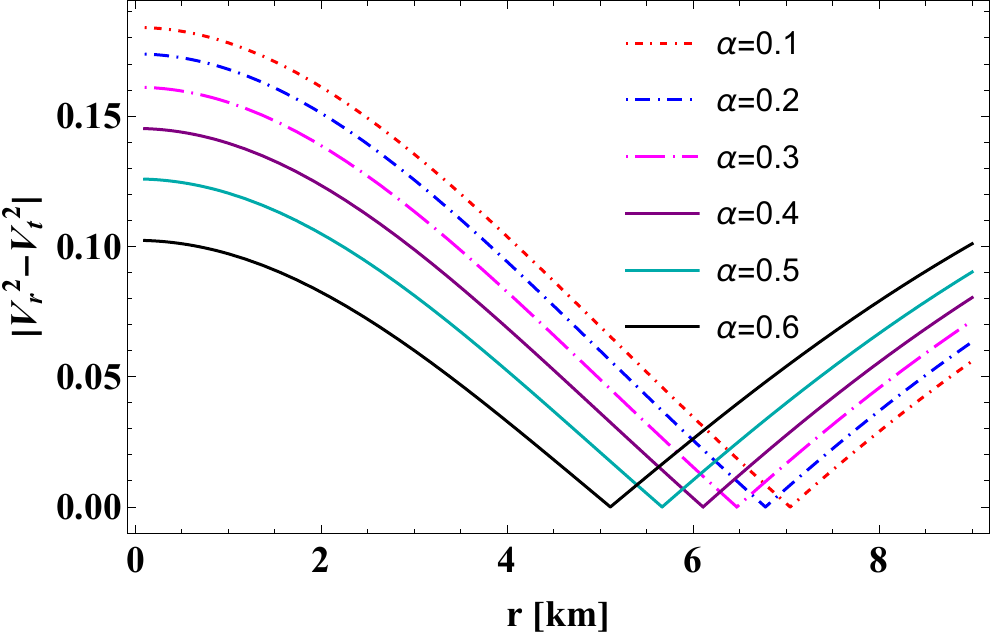}
    \caption{Stability analysis via speed of sound w.r.t. 'r' for $C = 0.288~ \text{km}^{-2}; D = 0.1; A = 0.009~ \text{km}^{-2}; B = 0.000009~\text{km}^{-4}; L = 0.0009 ~\text{km}^{-2}; N = 0.0009~ \text{km}^{-2}$.}
    \label{fig5}
\end{figure*}

The causality criterion and cracking method are the fundamental approaches to verifying the stability of stellar objects with the help of radial and transverse velocity of sounds. The causality criterion is one of the important conditions stated that the radial and transverse velocity of sounds should not exceed the speed of light, i.e., $v_r,\,v_t \leq 1$ (here speed of light is $1$ in the relativistic limit). For a stable star model, these velocities should be non-negative, i.e. $v_r,\,v_t \geq 0$. Combining these two conditions, we can write the following relations,
\begin{equation}
    0\leq v_r^2 \leq 1,\,\,\,\, 0\leq v_t^2 \leq 1.
\end{equation}
Where $v_r=\frac{dp_r^{\text{tot}}}{d\rho^{\text{tot}}}$ and $v_t=\frac{dp_t^{\text{tot}}}{d\rho^{\text{tot}}}$ are the radial and transverse velocity, respectively.
From Figure \ref{fig5}, we can observe that the causality conditions are satisfying for our anisotropic stellar dark star model. 
Moreover, Herrera's cracking method extends the causality criteria and ensures that the anisotropic compact stellar object is free from gravitational cracking and has stability \cite{her/1992, abr/2007}. To test this, we need to verify the following inequalities.
\begin{equation}
    |v_r^2-v_t^2|\leq 1.
\end{equation}
We present the above conditions graphically in Figure \ref{fig5}. One can easily observe that it follows Herrera's cracking condition. Moreover, one can see that as $\alpha$ approaches the highest value, the square of the speed of sound $v_r^2$ and $v_t^2$ tend towards its limit. From this point of view, we can make a conclusion that the higher value of the decoupling parameter leads to the instability of our constructed model.\\

\subsection{Harrison-Zeldovich-Novikov criterion}\label{secvc}

Perturbation analysis has always been one of the strong analyses when it comes to verifying the stability of any type of theory or cosmological model. Here, we shall discuss the condition for stability of an anisotropic stellar configuration under radial perturbation following the methodology demonstrated by Chandrasekhar in 1964 \cite{chan/1964}. For this analysis, we are going to use some of the perturbed physical quantities, such as the metric functions, energy density, and pressure, given by
\begin{equation}
    \lambda\rightarrow \lambda+\delta \lambda =ln[H(r)+\alpha\, \psi(r)]^{-1}-\frac{\delta H(r)+\alpha \,\delta \psi(r)}{H(r)+\alpha\, \phi(r)},
\end{equation}
\begin{equation}
    \nu\rightarrow \nu+\delta \nu = [G(r)+\alpha \, \phi(r)]+[\delta G(r)+\alpha \, \delta \phi(r)],
\end{equation}
\begin{equation}
    \rho\rightarrow \rho+\delta \rho,
\end{equation}
\begin{equation}
    p_r\rightarrow p_r+\delta p_r.
\end{equation}

Now, we are considering an oscillatory type of the radial perturbations, $e^{i\sigma t}$, where the stability of the oscillatory nature is determined by the characteristic frequency $\sigma$. Using the above-perturbed quantities, one can find the following relation from the first two field equations for the static spherically symmetric metric in the teleparallel equivalent general relativity (STGR) formulation,
\begin{multline}\label{71a}
    \sigma^2e^{\lambda-\nu}(p_r+\rho)\xi = \frac{d(\delta p_r}{dr}+\delta p_r \frac{d}{dr}\left(\frac{\lambda}{2}+\nu\right)\\
    +\frac{\delta \rho}{2}\frac{d\nu}{dr}-\frac{\rho+p_r}{2}\left(\frac{d\nu}{dr}+\frac{1}{r}\right)\left(\frac{d\lambda}{dr}+\frac{d\nu}{dr}\right) \xi.
\end{multline}
Where the Lagrangian displacement, $\xi$, is directly linked with the radial velocity $v=\partial \xi/\partial t$ in world time.

In addition, the perturbation of the energy density can be written as
\begin{equation}\label{72a}
    \delta \rho = -\frac{1}{r^2}\frac{\partial}{\partial r}[r^2 (p_r+\rho)\xi],
\end{equation}
and the conservation of the baryon number expressed as
\begin{eqnarray}\label{73a}
    \delta p_r =-\xi \frac{dp_r}{dr}-\Gamma_r\, p_r \frac{e^{\nu/2}}{r^2}\frac{d}{dr}[r^2 e^{-\nu/2\,\xi}],
\end{eqnarray}
here, $\Gamma_r$ is the radial adiabatic index.

Inserting  equations \eqref{72a} and \eqref{72a} in equation \eqref{71a} gives
\begin{multline}
    \sigma^2e^{\lambda-\nu}(p_r+\rho)\xi = \\ -\frac{d}{dr}\left(\xi \frac{dp_r}{dr}\right) 
    -\left(\frac{1}{2}\frac{d\lambda}{dr}+\frac{d\nu}{dr}\right) \xi \frac{dp_r}{dr}
    -\frac{\rho+p_r}{2}\left(\frac{d\nu}{dr}+\frac{1}{r}\right)\\ 
    \left(\frac{d\lambda}{dr}+\frac{d\nu}{dr}\right) -\frac{1}{2}\frac{d\nu}{dr}\left\lbrace\frac{d}{dr}[(p_r+\rho)\xi]+ \frac{2 (p_r+\rho)\xi}{r} \right\rbrace \\
    - e^{-(\lambda+2 \nu)/2}\frac{d}{dr}\left[e^{(\lambda+2 \nu)/2}  \frac{\Gamma_r p_r}{r^2} \, e^{\nu/2} \frac{d}{dr}(r^2e^{-\nu/2}\,\xi)\right],
\end{multline}
which can be simplified to
\begin{multline}\label{75}
    \sigma^2e^{\lambda-\nu}(p_r+\rho)\xi = \frac{4 \xi}{r}\frac{dp_r}{dr}-\frac{e^{-\lambda/2}}{e^{\nu}}\frac{d}{dr}\big[ e^{\frac{(\lambda+3\nu)}{2}}\, \frac{\Gamma_r\,p_r}{r^2}\\
    \frac{d}{dr}(r^2e^{-\nu/2}\,\xi)\big]+ 8\pi e^{\lambda} p_r (p_r+\rho)\xi-\frac{\xi}{p_r+\rho}\left(\frac{dp_r}{dr}\right).
\end{multline}
This equation is known as the \textit{pulsation equation}.

\begin{figure}
     \centering
    \includegraphics[height=6.2cm,width=7.8cm]{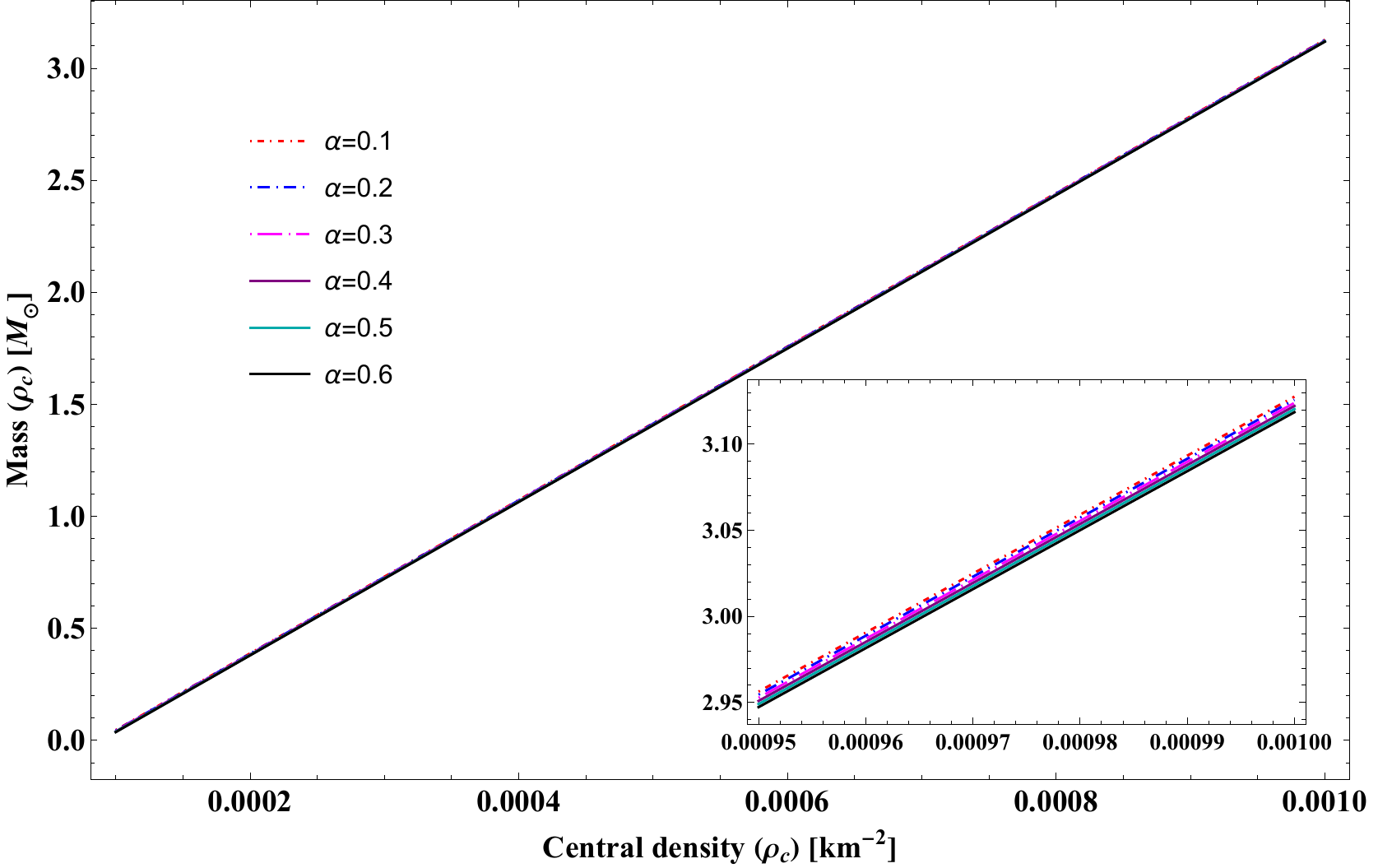}
    \caption{Graphical analysis of mass ($M_{\odot}$) w.r.t. central density $\rho_c$ for $C = 0.288~ \text{km}^{-2}; D = 0.1; A = 0.009~ \text{km}^{-2}; B = 0.000009~\text{km}^{-4}; L = 0.0009 ~\text{km}^{-2}; N = 0.0009~ \text{km}^{-2}$.}
    \label{fig6}
\end{figure}

Further, multiplying $r^2\xi e^{(\lambda+\nu)/2}$ in equation \eqref{75}, we get the \textit{characteristic equation} as
\begin{multline}
    \sigma^2 \int_0^R e^{(3\lambda-\nu)/2}(p_r+\rho)r^2\xi^2 \,dr = 4\int_0^R e^{(\lambda+\nu)/2}\frac{dp_r}{dr} \xi^2\,r dr\\
    -\int_0^R e^{\frac{(\lambda+3\nu)}{2}}\, \frac{\Gamma_r\,p_r}{r^2}
    \big[\frac{d}{dr}(r^2e^{-\nu/2}\,\xi)\big]^2 dr+ 8\pi \int_0^R e^{(3\lambda+\nu)/2} p_r\\
    (p_r+\rho)\xi^2 r^2 dr-\int_0^R  e^{(\lambda+\nu)/2}\frac{r^2\xi^2}{p_r+\rho}\left(\frac{dp_r}{dr}\right)^2 dr.
\end{multline}
The characteristic frequency $\sigma^2$ is found to be positive for small non-collapsing radial oscillations. Additionally, the polytropic equation of state $p_r = K\,\rho^{\Gamma_r}$ was utilized by Harrison et al. in 1965 \cite{her/1965} and Zeldovich, Novikov, and Silk in 1972 \cite{zel/1972} to simplify Chandrasekhar's calculations. Their analysis demonstrated that $\sigma^2$ remains positive if $\Gamma_r$ exceeds $\frac{4}{3}$. This leads to the mass being expressed as a function of central density, specifically $M(\rho_c) \propto \rho_c^{3(\Gamma_r - 4/3)/2}$. Notably, this implies that $M$ increases with $\rho_c$ when $\Gamma_r$ is greater than $\frac{4}{3}$, resulting in $dM / d\rho_c > 0$. This is called \textit{static stable criterion} or Harrison-Zeldovich-Novikov criterion. It states that the mass of an anisotropic stellar object must increase with respect to the central density for a stable configuration. 
To examine this criterion, we have the necessary physical quantities, which are expressed below,
\begin{eqnarray}\label{mass1}
   &&\hspace{0.1cm} \mathcal{M}(\rho_c)= \frac{1}{2} \Big(-\frac{1}{6A}B R^5 (\zeta_2+2 \alpha  L-16 \pi  \rho_c)-\frac{1}{6} R^3 (\zeta_2\nonumber\\ &&\hspace{1cm}+2 \alpha  L-16 \pi  \rho_c)+\frac{\zeta_2 R^3}{6}\Big)-\frac{1}{30} \alpha  L R^3 (3 N R^2-5),~~~~~\\ \label{mass2}
   &&\hspace{0cm} \frac{d\mathcal{M}}{d\rho_c}=\frac{1}{2} \left(\frac{8 \pi  B R^5}{3 A}+\frac{8 \pi  R^3}{3}\right).
\end{eqnarray}
For a stable stellar configuration, the total mass has to satisfy $\frac{d\mathcal{M}}{d\rho_c}>0$; otherwise, it can be said that the stellar configuration is unstable. This yields the relationship $A+B R^2>0$, which holds through our physical analysis since the parameters $A$ and $B$ are positive. Moreover, It can be seen that the total mass of the stellar object is increasing with the central density in Figure \ref{fig6}. In addition, it shows that the mass is positive and varies linearly with respect to central density throughout the stellar region. Therefore, the explored stellar dark star model satisfied this stability criterion. Also, it is observed that the decoupling constant has a lesser effect on the total mass.

\section{Measurement of maximum mass and radii of observed compact objects via M-R curves}\label{secvi}
\begin{figure*}
    \centering
    \includegraphics[height=6.2cm,width=8.6cm]{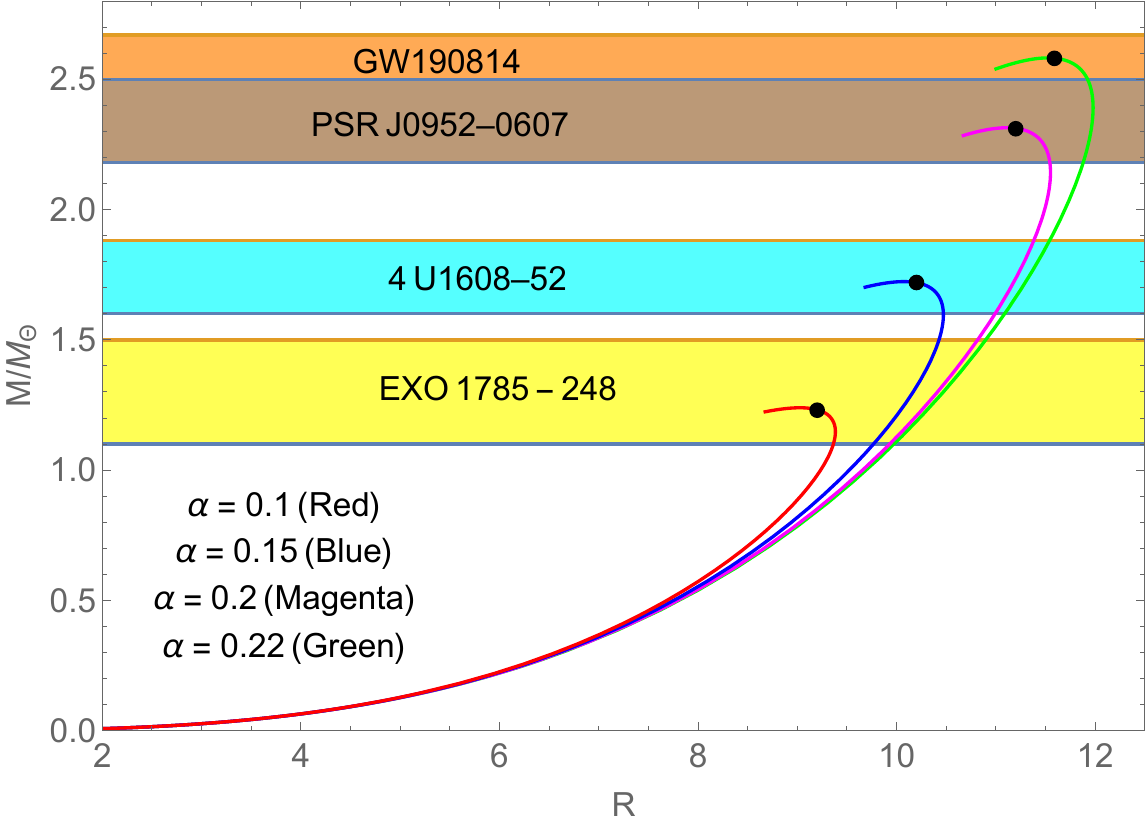}~~~
    \includegraphics[height=6.2cm,width=8.6cm]{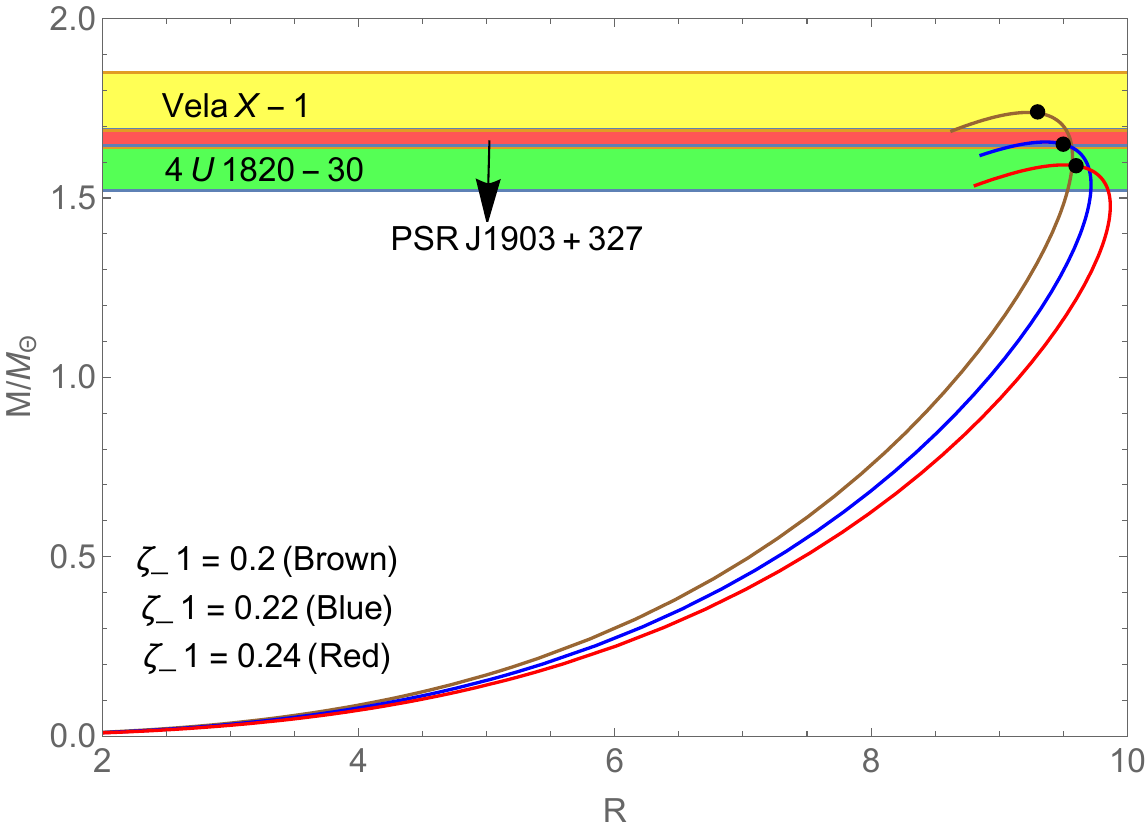}
        \caption{Prediction of radii of some well-known compact objects from our model for different values of the decoupling parameter $\alpha$ (left) and model parameter $\zeta_1$ (right). \label{mr}}
\end{figure*}
Furthermore, we generated the $M-R$ plot by varying the decoupling coupling parameter $\alpha$ in the left panel of Fig.~\ref{mr}. We notice that the associated radius is between $9.2$ and $11.59$ km, and the maximum masses range from $(1.23-2.58)M_{\odot}$ depending on the choices of the parameters $\zeta_1,\,\zeta_2,\,A,\,B,\,N$ and $L$. The $M-R$ curve enables us to predict the possible radius of some well-known compact stars. We observe that corresponding to the decoupling parameter $\alpha=0.10,\,0.15,\,0.20$, and $0.22$, $M-R$ curve encompasses a wider range of masses of compact stars like EXO 1785-248 [mass=$1.3_{-0.2}^{+0.2}~M_{\odot}$], 4U 1608-52 [mass=$1.74_{-0.14}^{+0.14}~M_{\odot}$], PSR J0952-0607 [mass=$2.35_{-0.17}^{+0.17}~M_{\odot}$], and the lighter component of the GW 190814 event [mass= $(2.50-2.67)~M_{\odot}$] respectively. It can be noticed that for the current model, the predicted radii have grown as the coupling parameter $\alpha$ increases. Bhar and Pretel \cite{Bhar:2023xku} recently reported the dark energy star model in the background of $f(Q)$ gravity and found a similar result. Through the analysis of the $M-R$ curve, the authors have demonstrated that as the coupling parameter $\alpha$ of $f(Q)$ gravity grows, the maximum mass and radius also increase. With a radius of $9.3$ kilometers, they arrive at a maximum mass of $2.57~M_{\odot}$. The parameter $\alpha$ dictating decoupling influences the degree to which dark matter (DM) impacts stellar characteristics. In our analysis, it is evident that the presence of dark matter within the fluid contributes to the formation of more massive stars. This outcome aligns with prior expectations as the interaction between dark matter and compact star matter intensifies, resulting in a stiffer equation of state (EOS). Such a stiffer EoS is capable of sustaining greater mass and higher compactness in stars.
Moreover, dark matter not only enhances the interactions with ordinary matter but also fortifies the interactions among ordinary matter particles themselves, acting as a cohesive substance akin to the plum pudding model. This cohesive, or 'gluing,' effect moderates the vibrations of ordinary matter caused by perturbations, thereby promoting a more stable stellar system. Table~\ref{tb1} displays the predicted radii for the above-mentioned compact stars for various values of $\alpha$. By analyzing the $M-R$ curve, the radii of EXO 1785-248, 4U1608-52, PSR J0952-0607, and the lighter component of the GW 190814 event are respectively obtained as $9.2,\,10.2,\,11.2$, and $11.59$ km. Based on selecting an appropriate parameter set, Table~\ref{tb1} indicates that our proposed model supports the mass of a compact star greater than $2~M_{\odot}$. Graphical analysis indicates that the maximum mass of our model is $2.58~M_{\odot}$ with an associated radius of $11.59$~km, corresponding to $\alpha=0.22$. This maximal mass lies in the mass gap range ($2.5-5$)~$M_{\odot}$. This finding satisfies the observational constraints and may be a candidate for the lighter objects in the GW 190814 event, as well as the large unseen companion in the binary system 2MASS J05215658+4359220, with mass $3.3^{+2.8}_{-0.7}~M_{\odot}$ \cite{Todd}.\par
On the other hand, in the right panel of Fig.~\ref{mr}, we have drawn the $M-R$ plots for three different choices of $\zeta_1$. The mass-radius curves cover the masses of the compact stars Vela X-1, 4U 1820-30, and PSR J1903+327, corresponding to $\zeta_1=0.2,\,0.22$, and $0.24$, respectively. One can note that when $\zeta_1$ grows, the maximal masses reduce, but the associated radii increase. So, in this case, the star becomes less massive as $\zeta_1$ increases. The predicted radii of the above-mentioned compact star from our model are presented in Table~\ref{tb2}.

\begin{table*}[t]
\centering
\caption{The prediction of radii for some well-known compact stars for different values of $\alpha$. }\label{tb1}
\begin{tabular}{@{}ccccccccccccc@{}}
\hline
$\alpha$& Observed mass && Predicted radius  && Matched with the mass of \\
& $M(M_{\odot})$ && (in km.) \\
\hline
0.10& 1.23 &&       9.2 && EXO 1745-248 \cite{Ozel:2008kb} \\
0.15& 1.72 &&       10.2 && 4U 1608-52 \cite{Rawls:2011jw} \\
0.20& 2.31 &&     11.2 && PSR J0952-0607 \cite{Romani:2022jhd}\\
0.22& 2.58 &&     11.59 && lighter component of \\
&&&&&GW 190814 event \cite{LIGOScientific:2020zkf}\\
\hline
\end{tabular}
\end{table*}

\begin{table*}[t]
\centering
\caption{The prediction of radii for some well-known compact stars for different values of $\zeta_1$. }\label{tb2}
\begin{tabular}{@{}ccccccccccccc@{}}
\hline
$\zeta_1$& Observed mass && Predicted radius  && Matched with the mass of \\
& $M(M_{\odot})$ && (in km.) \\
\hline
0.20& 1.74 &&       9.3 && Vela X-1 \cite{Rawls:2011jw} \\
0.22& 1.65 &&       9.5 && 4U 1820-30 \cite{Guver:2010td} \\
0.24& 1.59 &&     9.6 && PSR J1903+327 \cite{Freire:2010tf}\\
\hline
\end{tabular}
\end{table*}

\section{Mass measurement via equi-mass planes}\label{secvii}
\begin{figure*}
    \centering
    \includegraphics[height=6.2cm,width=8.6cm]{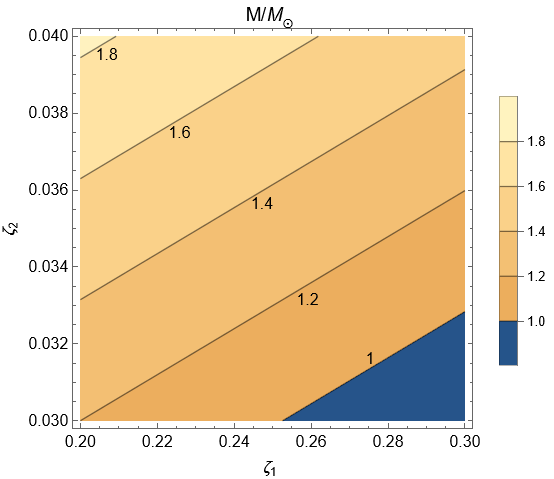}~~~
    \includegraphics[height=6.2cm,width=8.6cm]{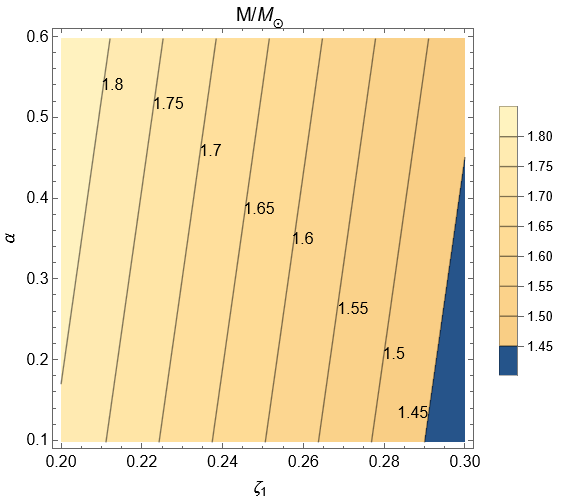}\\
    \includegraphics[height=6.2cm,width=8.6cm]{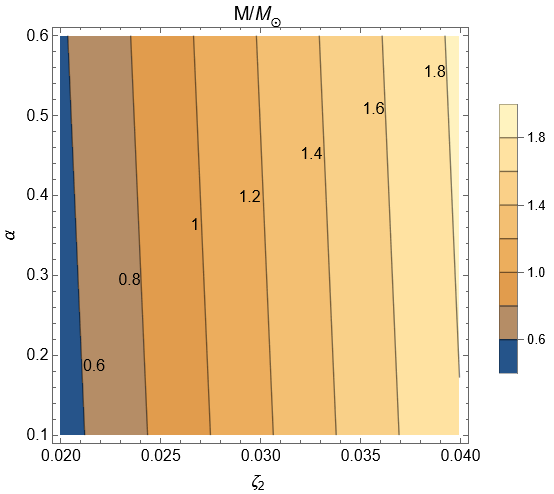}
    \caption{(Upper left panel) Equi-mass contour in $\zeta_1-\zeta_2$ plane, (upper right panel) in $\zeta_1-\alpha$ plane, and (bottom panel) in $\zeta_2-\alpha$ plane are shown. }
    \label{con}
\end{figure*}

In this section, a couple of contour graphs are drawn for an in-depth analysis of the mass of our present model. 
We have displayed the equi-mass contours in the $\zeta_1-\zeta_2$ and $\zeta_1-\alpha$ planes in the upper left and upper right panels, respectively of Fig.~\ref{con}. The upper left panel of Fig.~\ref{con} illustrates this point: the mass of the star decreases with increasing $\zeta_1$, whereas mass grows with increasing $\zeta_2$.
On the other hand, one can notice an increase in mass of the compact star with the growing value of $\alpha$ in the $\zeta_1-\alpha$ plane, as shown in the upper right panel of Fig.~\ref{con}. In contrast, this profile indicates that the mass reduces when $\zeta_1$ increases.\\
The equi-mass contour on the $\zeta_2-\alpha$ plane is displayed in the bottom panel of Fig.~\ref{con}. The profile shows that mass grows as $\zeta_2$ and the decoupling parameter $\alpha$ increases. So, the supermassive solution is desirable with higher values of $\zeta_2$ and a high decoupling parameter $\alpha$ as confirmed by the graphical analysis.

\section{Energy exchange between fluid distributions under dark matter}\label{secviii}
The study of self-gravitating systems is essential for comprehending the internal dynamics of compact stellar objects and the gravitational collapse process. These interiors are made up of several interacting fluids, making them challenging. Even if many fluids can be mixed to form a single energy-momentum tensor $T_{\mu\nu}^{tot}$, it is still difficult to figure out how they interact. This manuscript considers two arbitrary sources, $T_{\mu\nu}$ and $\Theta_{\mu\nu}$, for characterizing the fluid distribution inside the compact star. Therefore, the contracted Bianchi identities yields $\nabla_{\mu}T^{\mu\nu}+\nabla_{\mu}\Theta^{\mu\nu}=0$. There are two possibilities in this scenario. The first possibility is that each source is covariantly conserved. That can be expressed as
\begin{eqnarray}
    \nabla_{\mu}T^{\mu\nu}=0 ~~;\quad \nabla_{\mu}\Theta^{\mu\nu}=0.
\end{eqnarray}
The second one is a more intriguing and realistic option involving an energy exchange between these sources. This concept is mathematically represented as follows.
\begin{eqnarray}
    \nabla_{\mu}T^{\mu\nu}= -\nabla_{\mu}\Theta^{\mu\nu}.
\end{eqnarray}

\begin{figure*}
    \centering
    \includegraphics[height=6.2cm,width=8.5cm]{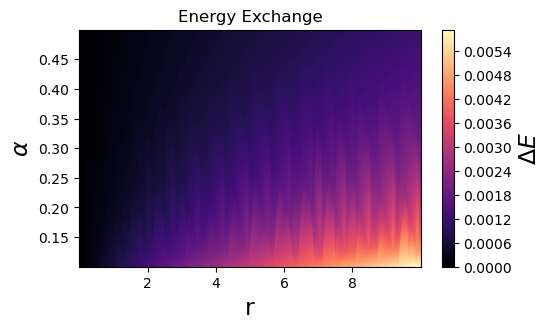}~~~
    \includegraphics[height=6.2cm,width=8.5cm]{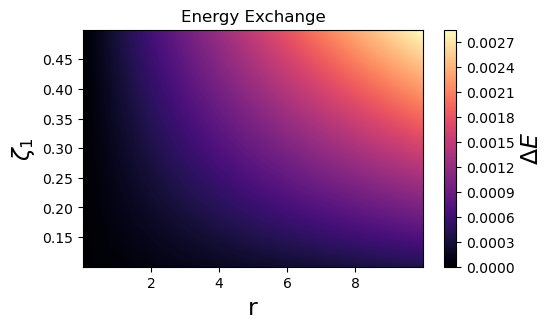}
    \caption{Energy exchange between the fluid and dark matter in the $r-\alpha$ plane (left) and $r-\zeta_1$ plane (right) for $C = 0.288~ \text{km}^{-2}; D = 0.1; A = 0.009~ \text{km}^{-2}; B = 0.000009~\text{km}^{-4}; L = 0.0009 ~\text{km}^{-2}; N = 0.0009~ \text{km}^{-2}$.}
    \label{enexpic}
\end{figure*}

In subsequent studies, authors \cite{enexc1} along with \cite{enexc2}, investigated a crucial aspect of energy exchange between two sources, denoted as \( T_{ij} \) and \( T^{\Theta}_{ij} \). Their calculations revealed that the energy transferred between these sources is represented by the symbol \( \delta E \) and is defined accordingly.
\begin{eqnarray}
    \delta E= \frac{\phi^{\prime}(r)}{2}(\rho+p_r).
\end{eqnarray}
Since for a stellar object, $\rho$ and $p_r$ are always positive in nature; therefore, it is quite obvious that the behavior of $\delta E$ will depend on $\phi^{\prime}(r)$. Hence, $\phi^{\prime}>0$ occurs $\delta E>0$,  in this case, the environment receives energy from the new source $T_{ij}^{\Theta}$. Conversely, when $\phi^{\prime}<0$, it returns to $\delta E<0$, and the new gravitational sector gets energy from the seed source. By, utilizing the Eq. (\ref{sfe1}), (\ref{sfe2}) and (\ref{phi}), we get the mathematical expression of the energy exchange between the dark matter source and the system environment is,
\begin{eqnarray}
    &&\hspace{0cm}\delta E = \frac{\phi_1}{2} \big(\zeta_1 \big(r^2 (2 A C+B)+A+2 B C r^4+2 C\big)\nonumber\\&&\hspace{1cm}-3 A \zeta_1-5 B \zeta_1 r^2 \big).
\end{eqnarray}
The variation of energy exchange throughout the star's region is illustrated in Fig.\ref{enexpic} for both in the $r-\alpha$ and $r-\zeta_1$ planes. For a specific decoupling parameter $\alpha$, it is evident that the energy exchange does not increase monotonically. Although it remains positive throughout the entire radius, there are regions where it increases and others where it decreases, resulting in an oscillatory behavior throughout the stellar region. Moreover, in the $r-\zeta_1$ plane, for a particular model parameter $\zeta_1$, the energy exchange consistently increases across the stellar radius. Additionally, it is observed that at the center of the star, the $\zeta_1$ parameter has minimal impact on energy exchange. However, as we approach the stellar boundary, the influence of this model parameter becomes more pronounced. Consequently, in both planes, it is evident that at the stellar surface, the energy exchange between the perfect fluid and dark matter is significantly greater.

\section{Moment of Inertia and Compactness}\label{secix}

The Bejger-Haensel \cite{moi1} method provides an approximation for the moment of inertia of a compact object by transforming a static system into a rotating one. The core concept involves treating the rotational effects as a perturbation of a static, spherically symmetric spacetime within a compact star. Recent estimates of the Crab Nebula's properties are utilized to derive constraints on the pulsar's moment of inertia, mass, and radius. To achieve this, we use an approximate formula that integrates these three parameters. The moment of inertia is approximated as follows $ I=a(x) M r^2;$. Where $a(x)$ can be defined qualitatively differently for neutron star (NS) and strange star (SS) as:
\begin{eqnarray}
    a_{NS}(x)=\frac{x}{0.1+2x}\quad \text{for}~~ x\leq 0.1,\\
    a_{NS}(x)=\frac{2}{9}(1+5x) \quad \text{for} ~~x> 0.1, \\
    a_{SS}(x)=\frac{2}{5}(1+x) \quad \text{for} ~~M>0.1 M_{\odot}.
\end{eqnarray}
In the above expression, the parameter $x$ is a dimensionless compactness parameter defined as $x=M/M_{\odot}(\text{km}/R)$.

\begin{figure*}
    \centering
    \includegraphics[height=6.2cm,width=8.5cm]{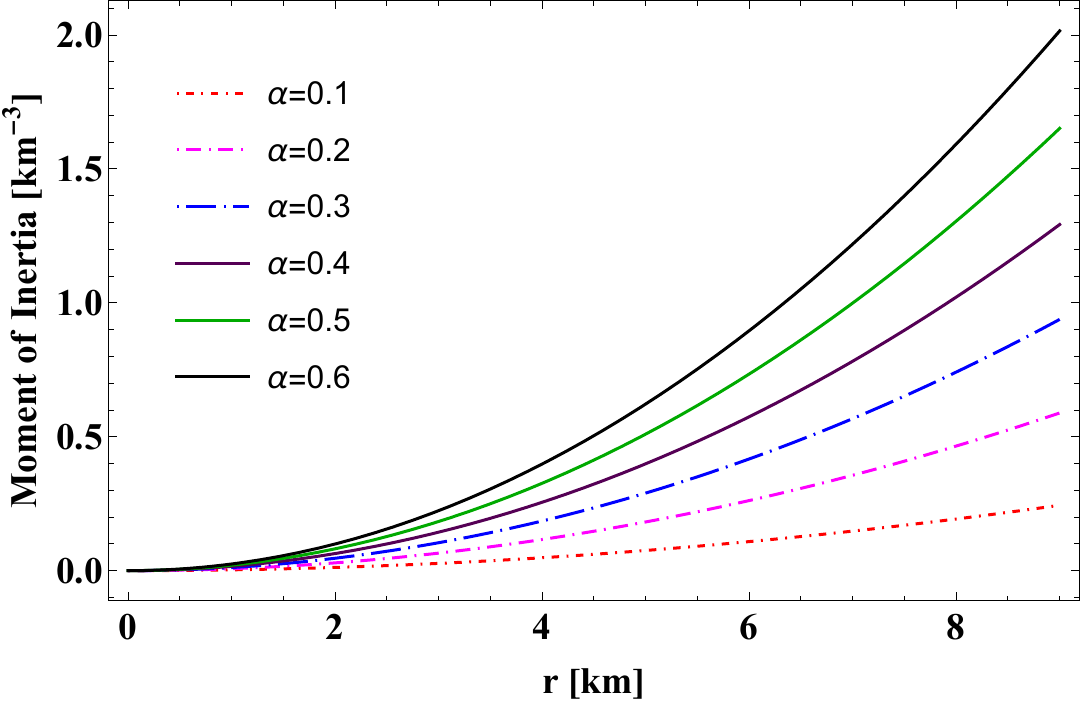}~~~
    \includegraphics[height=6.2cm,width=8.5cm]{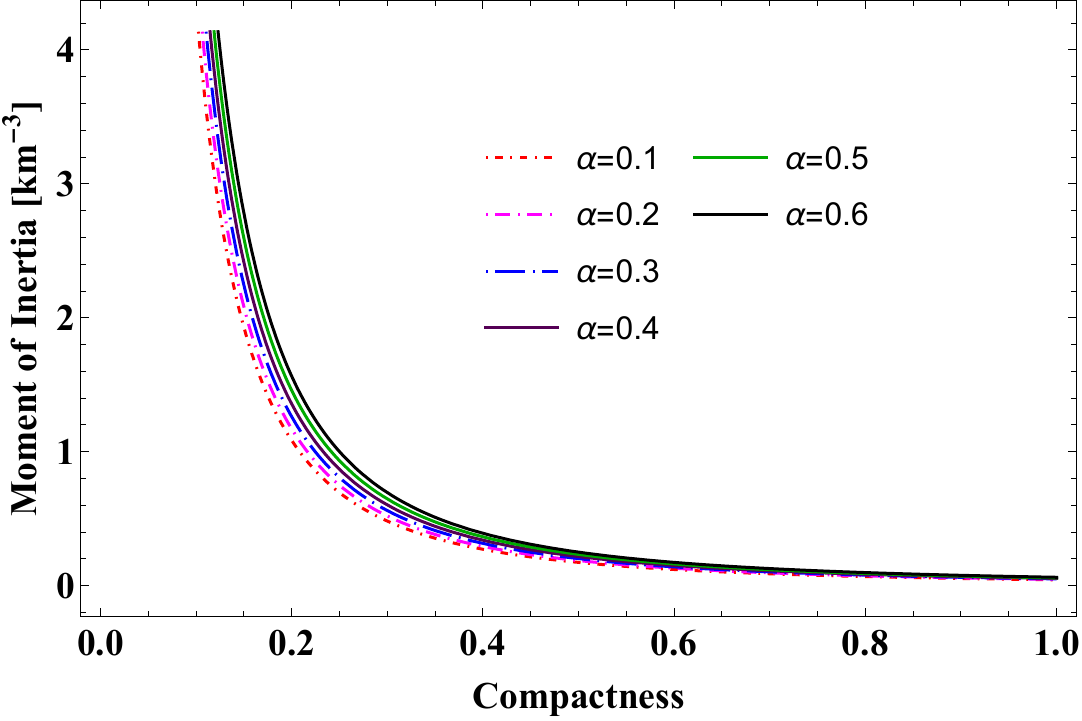}
    \includegraphics[height=6.2cm,width=8.5cm]{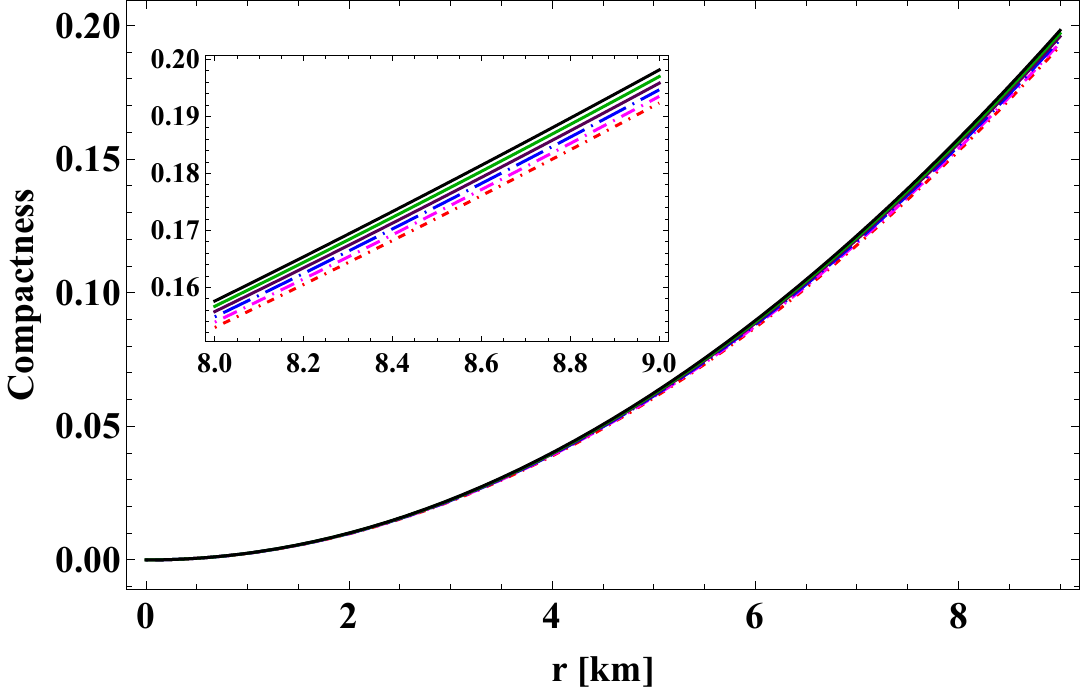}
    
    \caption{Profile of moment of inertia w.r.t. 'r' (upper-left) and moment of inertia w.r.t. compactness (upper-right) and compactness parameter w.r.t 'r' (bottom) for $C = 0.288~ \text{km}^{-2}; D = 0.1; A = 0.009~ \text{km}^{-2}; B = 0.000009~\text{km}^{-4}; L = 0.0009 ~\text{km}^{-2}; N = 0.0009~ \text{km}^{-2}$.}
    \label{moi}
\end{figure*}
Moreover, In our current model, the compactness factor is defined as \( u(r) = \frac{\mathbf{m}(r)}{r} \). This factor is crucial for classifying compact objects into the following categories: (i) regular stars (\( u \sim 10^{-5} \)), (ii) white dwarfs (\( u \sim 10^{-3} \)), (iii) neutron stars (\( 0.1 < u < 0.25 \)), (iv) ultra-compact stars (\( 0.25 < u < 0.5 \)), and (v) black holes (\( u = 0.5 \)).

The variation in the moment of inertia and the compactness factor is illustrated in Fig.\ref{moi}. It is evident that the moment of inertia increases monotonically with respect to the stellar radius, reaching its peak value at the star's surface. Moreover, it has been observed that when decoupling parameter 
$\alpha$ increases and the moment of inertia rises significantly. Apart from that, the second panel in Fig.\ref{moi} depicted the physical behavior of the moment of inertia w.r.t the compactness parameter. The compactness and moment of inertia of a star are closely linked through the star's mass distribution and density profile. Higher compactness generally leads to a lower moment of inertia, reflecting the concentration of mass towards the center, which can be noticed in Fig.\ref{moi}(b). This relationship is crucial for understanding the rotational dynamics and internal structure of various types of stars.

Furthermore, Fig.\ref{moi} (bottom panel) depicts the compactness profile for our model, which also shows a monotonically increasing function of
$r$ and remains within the range of $(0, 0.20)$. Consequently, this allows us to classify our constructed compact star model within the neutron star (NS) category.

\section{A comparative study of stellar model}\label{sec10}

In this section, we will provide a succinct overview of recent advancements in the study of compact stars through a comparative study. In a recent study by Maurya \cite{maurya}, the author explored the vanishing complexity factor within the framework of $f(Q)$ gravity, employing the Vlasenko-Pronin spacetime. The findings indicated that neutron stars exhibit higher compactness with an increase in the decoupling parameter, akin to the results of our study. Moreover, their research discovered that the presence of dark matter within the fluid contributes to the formation of more massive stars, corroborating our findings. This outcome conforms to previous expectations, as the interaction between dark matter and compact star matter becomes more pronounced, resulting in a stiffer EoS. Such a stiffer EoS is capable of supporting a greater mass and increased compactness in stars.

The research in \cite{rv1} examines compact stars within the framework of \(f(R)\) gravity, utilizing scalar-curvature modifications to explore mass-radius relations and address the mass gap region. Our study, in comparison, applies teleparallel \(f(T)\) gravity and the Complete Geometric Deformation (CGD) method to investigate the role of torsion, gravitational decoupling, and dark matter effects. While both works aim to interpret phenomena like the GW190814 event, \(f(R)\) gravity focuses on altering the curvature scalar, whereas \(f(T)\) gravity modifies the torsion scalar. Moreover, our use of a strange quark equation of state combined with the decoupling parameter \(\alpha\) offers a distinct pathway to modeling masses in the gap region. Together, these approaches highlight the versatility of modified gravity theories in enriching the understanding of compact stars.

To further compare with $f(Q)$ gravity, it is imperative to highlight the research by Bhar and Pretel \cite{PB2}, which demonstrated that both the maximum mass and the corresponding radius of dark energy stars augment as the coupling parameter in $f(Q)$ gravity increases. Conversely, in the context of $f(Q,T)$ gravity, the findings suggest that a decrease in the coupling parameter between non-metricity and matter leads to the attainment of the maximum mass limit for a compact object. Additionally, in the realm of quadratic $f(Q)$ gravity, the presence of a quintessence field results in an increase in the maximum mass and corresponding radius as the coupling parameter value decreases \cite{106}.

Moreover, other studies have undertaken a meticulous analysis of the maximum permissible mass in compact objects using the $M-R$ curve within the modified gravity framework \cite{107,108}. In another study \cite{shym}, by comparing the results with a slowly rotating configuration, researchers have also examined the moment of inertia and the period of rotation employing the Bejger-Haensel approach. They found that the maximum inertia was achieved for a mass of $3.106~M_{\odot}$.

The Occam's razor principle prioritizes simplicity, often favouring explanations that require the fewest assumptions. The GW190814 event suggests that the secondary object in the mass gap is likely a low-mass black hole (BH) or neutron star (NS) within the framework of GR, as discussed in \cite{rv2}. Our study takes a different approach by combining a strange quark EoS with teleparallel gravity through the CGD method. Although this introduces additional complexity, it enables us to model compact stars in the mass gap by modifying the TOV equations. With specific parameter choices such as \(\zeta_1 = 0.9\), \(\zeta_2 = 0.001\), and varying \(\alpha\), our results align with the observed mass-radius relations for GW190814. This demonstrates the potential of alternative gravity theories to address observational challenges while complementing simpler explanations.

\section{Conclusion}\label{secx}
In this study, we have developed a geometrically deformed strange star within the framework of teleparallel gravity by applying GD for the first time. The key finding of this work is the successful derivation of an exact solution for deformed SS models utilizing the CGD technique. In this investigation, we have considered the scenario of a vanishing complexity factor, which has introduced a novel avenue for generating solutions to the field equations in teleparallel gravity for the spherically symmetric structure of celestial bodies. By employing the feasible solution (Tolman-Kuchowickz)\cite{tolman,kucho} on the system of two metric potentials, namely $g_{tt}$ and $g_{rr}$, we have solved the seed system of our constructed model. Additionally, we consider that space-time is deformed by the presence of cold dark matter (CDM) within DM halos, leading to perturbations in the $g_{tt}$ and $g_{rr}$ metric potentials where the decoupling parameter dictates the extent of DM content. This DM deformation is executed mathematically for solving the $\Theta$ gravitational sector in CGD method. The nature of dark matter, whether cold, warm, or interacting, significantly influences its astrophysical signatures and interactions within compact objects. For instance, the effects of collisional dark matter on compact stars, as discussed in \cite{a1}, highlight potential alterations in stellar dynamics and evolution. The authors propose a potential method for detecting the effects of the particle nature of dark matter using observational data from the Guitar Nebula. This approach is applicable if dark matter is collisional, involving interacting particles. Without committing to a specific model for interacting dark matter, the authors adopt an agnostic perspective, hypothesizing that the Guitar Nebula bow shock is produced by the interaction of a high-speed neutron star traversing an interstellar medium composed of both interacting dark matter and hydrogen. Similarly, reviews such as \cite{a2} emphasize the astrophysical relevance of warm dark matter, particularly in resolving discrepancies at smaller galactic scales.
\\ The significant features of our findings are given below:

\begin{enumerate}
    \item \textbf{Nature of physical quantities :} The physical quantities $\rho^{tot}$, $p_r^{tot}$, and $p_t^{tot}$ vary with changes in the decoupling parameter $\alpha$, as shown in Figures \ref{fig1} and \ref{fig2}. These figures indicate that as $\alpha$ increases, the energy density throughout the anisotropic dark star also rises, suggesting that the new source term $\Theta_{\mu\nu}$ leads to the formation of denser stellar objects. Figure \ref{fig1} shows that the radial pressure decreases with higher $\alpha$ in the central region, converges near the surface, and drops to zero around 9 km. Similarly, Figure \ref{fig2} illustrates that the tangential pressure converges to different finite, non-zero values at the surface for each $\alpha$. This convergence indicates that the surface pressures, both radial and tangential, remain unaffected by changes in $\alpha$, producing effects similar to those seen in isotropic systems. We graphically present the behavior of the gradients in Figure \ref{fig4}. One can easily observe that the gradients satisfy the conditions $\frac{d\rho^{\text{tot}}}{dr}<0,\,\,\,\frac{dp_r^{\text{tot}}}{dr}<0,\,\,\,\frac{dp_t^{\text{tot}}}{dr}<0.$ as well as they vanish at $r=0$  which is one of the crucial aspect for a stable stellar configuration. Moreover, our constructed model satisfies the energy conditions (NEC, DEC, SEC, WEC) proving that no exotic matter is present inside the constructed SS model.
    \item \textbf{Discussion on stability analysis:}
    We have verified the stability of our constructed model using the adiabatic index, Herrera's cracking method, and the Harrison-Zeldovich-Novikov criterion. Both the adiabatic index analysis and Herrera's cracking method demonstrate that small values of the decoupling parameter $\alpha$ can replicate the stability of the anisotropic dark star model. As shown in Figure \ref{fig3}, the adiabatic index ($\Gamma^{\text{tot}}$) remains greater than $4/3$ for various values of $\alpha$. Additionally, from Figure \ref{fig2} and Figure \ref{fig3}, we observe that while $\alpha$ has minimal impact on $\Delta^{\text{tot}}$ near the star's center, it significantly affects $\Gamma^{\text{tot}}$ for small radii. Small values of $\alpha$ thus ensure stability near the center, with no impact at the surface where $\Gamma^{\text{tot}}>4/3$ for all positive $\alpha$.

Herrera's cracking condition, $|v_r^2-v_t^2| \leq 1$, is graphically confirmed in Figure \ref{fig5} for all values of $\alpha$. As $\alpha$ approaches higher values, $v_r^2$ and $v_t^2$ approach the limit of 1, indicating potential instability at higher $\alpha$ values.

Regarding the Harrison-Zeldovich-Novikov criterion, for a stable stellar configuration, the total mass must satisfy $\frac{dM}{d\rho_c}>0$, implying $A + B R^2 > 0$, which our analysis confirms as $A$ and $B$ are positive. Figure \ref{fig6} illustrates that the total mass of the stellar object increases with central density, remaining positive and linearly varying with central density throughout the stellar region. The decoupling constant $\alpha$ has minimal effect on the total mass, reinforcing that our stellar dark star model satisfies this stability criterion.

\item \textbf{Maximum mass measurement :} The radii of a few compact star candidates were predicted using the $M-R$ curves that resulted from our models by varying the two parameters, $\alpha$ and $\zeta_1$, that are displayed in Tables~\ref{tb1}-\ref{tb2}. The profiles in Fig.~\ref{mr}(left) show that the maximum permissible masses and the related radii rise with an increase in the value of the decoupling parameter $\alpha$. We find that the radius of the compact stars lies between $9.2$ km and $11.59$ km, with the maximum masses ranging from $(1.23 - 2.58) M_{\odot}$, depending on the values of the parameters $\zeta_1$, $\zeta_2$, $A$, $B$, $N$, and $L$. The $M$-$R$ curve provides insights into the possible radii of well-known compact stars. For decoupling parameters $\alpha = 0.10$, $0.15$, $0.20$, and $0.22$, the $M$-$R$ curve accommodates a broad mass range for stars like EXO 1785-248 ($1.3_{-0.2}^{+0.2} M_{\odot}$), 4U 1608-52 ($1.74_{-0.14}^{+0.14} M_{\odot}$), PSR J0952-0607 ($2.35_{-0.17}^{+0.17} M_{\odot}$), and the lighter component of GW 190814 ($2.50 - 2.67 M_{\odot}$)respectively. Thus, in this modified theory of gravity, it will be more convenient to use a larger decoupling parameter $\alpha$ to describe higher-mass compact stars. The parameter $\alpha$, which governs the decoupling process, plays a crucial role in determining how dark matter affects the properties of stars. Our analysis reveals that the inclusion of dark matter within the stellar fluid leads to the creation of more massive stars. This finding is consistent with previous theoretical predictions. The $M-R$ curve in Fig.~\ref{mr}(right) shows that $M_{max}$ increases but the radii decrease as the model parameter $\zeta_1$ decreases, which implies that for a lower value of $\zeta_1$, the star becomes more compact.

\item \textbf{Energy exchange between two sources:} The variation of energy exchange between fluid distribution under dark matter throughout the star's region is illustrated in Fig.(\ref{enexpic}) for both in the $r-\alpha$ and $r-\zeta_1$ planes. In the $r-\alpha$ plane, when considering a specific decoupling parameter $\alpha$, the energy exchange doesn't follow a monotonic trend. Instead, it is increasing oscillatory in some regions and decreasing in others across the entire stellar radius. This behavior highlights the intricate interplay between the perfect fluid and dark matter. In the $r-\zeta_1$ plane for a fixed model parameter $\zeta_1$, the energy exchange consistently increases as we move outward from the star's center. Interestingly, at the core of the star, the $\zeta_1$ parameter has minimal impact on energy exchange. However, as we approach the stellar surface, its influence becomes more pronounced. Notably, at the stellar surface, the energy exchange between the perfect fluid and dark matter is significantly higher.

\item \textbf{Moment of inertia and compactness:}
Finally, we conducted an in-depth physical analysis of the moment of inertia relative to the compactness parameter using the Bejger-Haensel method \cite{moi1}. This relationship is pivotal for comprehending the rotational dynamics and internal structure of various types of stars. The core concept entails treating rotational effects as a perturbation of a static, spherically symmetric spacetime within a compact star. The variation in the moment of inertia and the compactness factor is depicted in Fig.\ref{moi}. It is apparent that the moment of inertia increases monotonically with the stellar radius, reaching its maximum value at the star's surface. Additionally, we observed that an increase in the decoupling parameter $\alpha$ results in a significant rise in the moment of inertia.\\
\indent Moreover, the second panel in Fig.\ref{moi} illustrates the moment of inertia's behavior concerning the compactness parameter. The compactness and moment of inertia of a star are intricately linked through the star's mass distribution and density profile. Higher compactness generally leads to a lower moment of inertia, reflecting the concentration of mass towards the center, as shown in Fig.\ref{moi}(b). Furthermore, the bottom panel of Fig.\ref{moi} presents the compactness profile for our model, which is a monotonically increasing function of $r$ and remains within the range of $(0,0.20)$. This classification indicates that our constructed compact star model falls within the neutron star (NS) category.
\end{enumerate}
Thus, it can be deduced that our model satisfies numerous physical attributes and stability criteria essential for a physically plausible anisotropic compact configuration. Therefore, our findings shed light on the complex dynamics within anisotropic dark stars. It built profound insights into the intricate relationship between anisotropy and the gravitational effects in teleparallel gravity, thereby advancing our comprehension of strange stars and their underlying physical mechanisms. Consequently, this model holds significant astrophysical relevance for the investigation of anisotropic compact stars.

\section*{Data availability} No new data was generated or analyzed to support this investigation.

\section*{Acknowledgement}
SP \& PKS  express gratitude to the National Board for Higher Mathematics (NBHM) under the Department of Atomic Energy (DAE), Government. of India, for providing financial assistance to conduct the Research project No.: 02011/3/2022 NBHM(R.P.)/R \& D II/2152 Dt.14.02.2022. P.B. is thankful to the Inter-University Centre for Astronomy and Astrophysics (IUCAA), Pune, Government of India, for providing visiting associateship. SM acknowledges the Japan Society for the Promotion of Science (JSPS) for providing a postdoctoral fellowship during 2024-2026 (JSPS ID No.: P24026). This work of SM is supported by the JSPS KAKENHI Grant (Number: 24KF0100). The work of KB was partially supported by the JSPS KAKENHI Grant Number 21K03547 and 23KF0008.

\end{document}